\documentclass[aps,prd,preprint,amsmath,citeautoscript,longbibliography,nofootinbib]{revtex4-2} 
\usepackage[utf8]{inputenc}
\usepackage{bm}
\usepackage{amsmath,mathrsfs,amsfonts}
\usepackage{graphicx,afterpage}
\usepackage{subcaption}
\usepackage{amsmath,latexsym}
\usepackage{color,soul}
\usepackage{float}
\usepackage[section]{placeins}
\usepackage{silence}
\begin{document}
\title{New Rotating Black Hole Solutions With Imperfect Fluid Energy-Momentum Tensor In $f(R)$ Gravity}
\author{Bardia H. Fahim{\footnote{bardia.fahim@usask.ca}} and A. M. Ghezelbash{\footnote{masoud.ghezelbash@usask.ca}}}

\affiliation{Department of Physics and Engineering Physics, University of Saskatchewan, Saskatoon SK S7N 5E2, Canada}
\date{\today}

\begin{abstract}
We find a new class of exact solutions for rotating black holes in  $f(R)$ gravity in presence of imperfect fluid. We find that the exact solutions are holographically dual to a hidden conformal field theory.  Moreover we consider another class of exact solutions for rotating black holes in  $f(R)$ gravity, which is obtained through a Lorentz boost transformation on a static spherical symmetry solutions in $f(R)$ gravity.  We also find that the {\color{black} exact boosted} solutions are holographically dual to a hidden conformal field theory. Comparing the CFT temperatures and the mode numbers of the two CFTs, we suggest a {\it conjecture} that, 
{\it {``Two identical rotating black holes (the same mass and the horizon) in $f(R)$ theory could be distinguished by looking at their dual CFT temperatures, and the dual CFT mode numbers.  The rotating black hole in the presence of matter has higher CFT temperatures and higher CFT mode numbers, compared to the vacuum boosted rotating black hole".}}
\end{abstract}

\maketitle
\newpage
\section{Introduction}
\label{sec:intro}

The holographic correspondence between a rotating black hole and a dual Conformal Field Theory (CFT) was the subject of intense research during the last fifteen years. The research was focused mainly to establish the correspondence for the four and higher dimensional extremal, as well as non-extremal black holes in the Einstein gravity coupled to the matter fields. The first known example of the holography, known as the Kerr/CFT correspondence, was shown explicitly in  \cite{stro}. The authors showed that the physics of the extremal Kerr black holes can be read from a dual CFT.  The physical results for an extremal Kerr black holes, such as the Bekenstein-Hawking entropy and the super-radiant modes of the black hole, 
are shown to be in agreement with the dual results in CFT.  In fact, the diffeomorphisms of the near-horizon geometry for the extremal Kerr black holes produce the generators of the conformal symmetry. 

The holographic correspondence also works for other different four and higher dimensional extremal rotating black holes in Einstein gravity \cite{KerrCFT1}-\cite{SaktiAnnPhys2020}.  In fact, for all the different extremal rotating black holes,  the near-horizon metric has a sub-metric which describes an exact Anti de Sitter (AdS) space.  The underlying isometries of the AdS can be enhanced to make the generators of a Virasoro algebra, which in turn makes a dual CFT description for the extremal black holes.

On the other hand, the lack of AdS structure in the near-horizon geometry of the non-extremal rotating black holes, hinders the enhancement of the AdS isometries to a Virasoro algebra, hence the situation becomes more complicated.
For the non-extremal rotating black holes, the authors of  \cite{cms} showed that instead of usual enhancement procedure for the extremal rotating black holes, we can find the dual conformal symmetry (known as the hidden conformal symmetry). In fact, for the non-extremal rotating black holes, the conformal symmetry is encoded in the solution space of a probe field in the background of non-extremal rotating black holes. The existence of the hidden conformal symmetry  was established for many non-extremal rotating black holes in four and higher dimensional Einstein gravity \cite{othe}-\cite{KSChen}.

In the context of the duality, the authors of \cite{othe}-\cite{KSChen} show that the macroscopic quantities for the non-extremal rotating black holes solutions of General Relativity (GR), can be obtained from the 
corresponding microscopic quantities of the CFT. For example, there is a perfect match between the macroscopic Bekenstein-Hawking entropy of the rotating GR black holes and the Cardy entropy of the CFT.
As an another example, the 
super-radiant bulk scatterings off the rotating black holes are in precise agreement with the scattering amplitudes of the CFT. 

In this article, we consider two different types of rotating black holes in the context of  modified theories of gravity  (MTG) \cite{25}-\cite{OLD6},  and in particular, the $f(R)$ gravity. 
The MTG  are theories to address the issues of GR, such as dark energy, since the accelerating expansion of the universe was discovered. We also note that in the appropriate limits, the MTG should produce the results that are in agreement with GR. In the modified theory of $f(R)$ gravity, the action includes a function of Ricci scalar $R$, which leads to a new class of MTG  {\cite{25}-\cite{N1}.

In this article, inspired with the existence of the dual CFT to the rotating black holes of GR, we consider two different rotating black holes in $f(R)$ gravity. We are interested to investigate the possible existence of the black hole holography in the class of $f(R)$ gravity, because in these theories the action is a non-trivial function of the Ricci scalar. The very different field equations may hinder the existence of the duality between the rotating black holes of $f(R)$ and a CFT. In this article, we focus only on the inverse $f(R)$ gravity, where $f(R)$ includes a term proportional to the inverse of the Ricci scalar. For this class of $f(R)$ theories, we consider two types of black hole solution. In type I, we use the {\color{black} Synge} G method to construct a rotating black holes in the background of a non-trivial energy-momentum tensor \cite{GAR}. We numerically show that the energy-momentum tensor satisfies all the necessary energy conditions, as it is impossible to show them analytically. In type II, we use a Lorentz boost transformation, to find a rotating black hole solution from a static black hole solution. Based on our CFT results, we come up with a conjecture about the two different types of black hole solutions in $f(R)$ gravity. 

We organize the article as follows. In section \ref{sec:sec2scalarfield}, we briefly review the $f(R)$ theory and its field equations and present the static black hole solutions. In section \ref{sec:sec3}, we explore and find a new rotating black hole solution in $f(R)$ theory, in presence of a non-trivial imperfect fluid.  We also establish the existence of a CFT dual to the type I rotating black holes.  

In section \ref{sec:sec4}, we apply a Lorentz boost transformation on the static black hole (in section \ref{sec:sec2scalarfield}) to find a ``boosted'' rotating black hole. We also establish the existence of a dual CFT to the rotating black holes. In section \ref{sec:sec5}, we compare our CFT results for type I and type II rotating black holes and  present a conjecture about the black holes of $f(R)$ theory.
We wrap up the article by concluding remarks in section \ref{sec:conc}, and two appendices.


\section{Static black hole solutions in $f(R)$ gravity }\label{sec:sec2scalarfield}

 The action of $f(R)$ gravity in 4-dimensions is represented by
\begin{equation}
    I=\frac{1}{16\pi G}\int d^4x\sqrt{-g}(f({R})+\mathcal{L}^{(matter)}),\label{action}
\end{equation}
where $f(R)$ is a function of the Ricci scalar, $\mathcal{L}^{(matter)}$ is the Lagrangian of matter fields and $G$ is the gravitational constant.  In equation \eqref{action} and through the article,  we use the geometrized units such that $G=c=1$. We vary the action (\ref{action}) with respect to the metric $g_{\mu \nu}$, and find the field equations of $f(R)$ gravity as
\begin{equation}
    R_{\mu\nu}f_{{R}}-\frac{1}{2}g_{\mu\nu}f(R)-\nabla_{\mu}\nabla_{\nu}f_{{R}}+g_{\mu\nu} \Box f_{{R}}=T_{\mu\nu}^{(matter)}, \label{e2}
\end{equation}
where $R_{\mu\nu}$ is the Ricci tensor, $f_{{R}}\equiv \frac{df({R})}{d{R}}$, $\Box$ is D'Alembertian operator $\Box\equiv\nabla_{\alpha}\nabla^{\alpha}$, and $T_{\mu\nu}^{(matter)}$ is the energy-momentum tensor for matter fields, which is derived from $\mathcal{L}^{(matter)}$. 

We focus on a well-known static spherically symmetric solutions to the vacuum $f(R)$ gravity, that satisfies the field equations (\ref{e2}), with the following line element
\begin{equation}
    ds^2=-g(r)dt^2+\frac{1}{g(r)}dr^2+r^2d\Omega^2_{(2)}. \label{e3}
\end{equation}
The solution for 
\begin{equation}
f(R)=R-\frac{16\Lambda^2}{3R}, \label{fr}
\end{equation}
where $\Lambda$ is a constant, 
is given by the metric function \begin{equation}
    g(r)=1-\frac{M}{r}-\frac{\Lambda}{3}r^2, \label{e10}
\end{equation}
where $M$ and $\Lambda$ are related to the black hole mass and the cosmological constant, respectively. We notice that for the large values of $R$, we can neglect the second term in equation (\ref{fr}), and so, we have $f(R)=R$, which leads to the theory of general relativity. However, for the small values of the Ricci scalar, we get non-trivial modified gravity. It is worth noting that, we started with a $f(R)$ model without the cosmological constant term $\Lambda g_{\mu\nu}$ in the field equations (\ref{e2}) (or no constant term in the action (\ref{action})), in analogy with the general relativity. However due to the presence of term $\frac{16\Lambda^2}{3R}$ in $f(R)$, we can interpret the field equations (\ref{e2}), as the Einstein field equations, in the presence of the cosmological constant. In fact, this may suggest that the cosmological constant emerges from the $f(R)$ theory of gravity \cite{hendi}.

In this article, we consider two very different types of black hole solutions to $f(R)$ gravity with equation (\ref{fr}). In type I black hole solution, we use a metric which is locally rotating, with an inherent rotational parameter in the presence of a non-trivial imperfect fluid. In type II black hole solution, we simply apply a Lorentz boost transformation to the static metric (\ref{e3}) of vacuum $f(R)$ theory, to construct a globally rotating black hole solution. As we show in sections \ref{sec:sec3} and \ref{sec:sec4}, we explicitly  construct CFT duals to both black holes in $f(R)$ gravity. We then discuss and compare the results of the dual CFTs, and come up with a conjecture that enable us to predict the nature of rotation in the two black holes, merely from the dual CFT results. We stress that rotation in type I black hole, is inherent to the black hole, while in type II black hole, is merely a Lorentz boost. In other words, in the latter case, one can find a reference frame in which the observer would co-rotate with the black hole. Such frame does not exist for type I black hole.

\section{Type I black hole solutions in $f(R)$ gravity}\label{sec:sec3}

\subsection{New rotating black hole solutions in $f(R)$ gravity}
We adopt a modification of the Newman-Janis algorithm (NJA) proposed by Azreg-Aïnou to generate a rotating line element given by \cite{azreg2014static}
\begin{eqnarray}
       ds^2 &=& -\frac{(F(r)H(r)+a^2\cos^2{\theta})\psi}{(K(r)+a^2\cos^2{\theta})^2}dt^2+\frac{\psi}{F(r)H(r)+a^2}dr^2- 2a\sin^2{\theta}\Bigl(\frac{K(r)-F(r)H(r)}{(K(r)+a^2\cos^2{\theta})^2}\Bigl)\psi dtd\phi
       \nonumber \\
       &+& \psi d\theta^2  +\psi\sin^2{\theta}\Bigl[1+a^2\sin^2{\theta}\Bigl(\frac{2K(r)-F(r)H(r)+a^2\cos^2{\theta}}{(K(r)+a^2\cos^2{\theta})^2}\Bigl)\Bigl]d\phi^2, \label{AA metric}
\end{eqnarray}
with $a$ the rotation parameter. In the line element (\ref{AA metric}), $F(r)$, $G(r)$ and $H(r)$ are metric functions, and $K(r)$ and $\psi$ are given by $K(r)=H(r)\sqrt{\frac{F(r)}{G(r)}}$ and $\psi=K(r)+a^2\cos^2{\theta}$. We should note that we do not need the complexification step, which exists in the NJA. Moreover, unlike the NJA, the metric (\ref{AA metric}) can be written in the Boyer-Lindquist coordinates \cite{pal2023rotating}. We also note that in the limit of $a=0$, we find the  rotating metric (\ref{AA metric})  reduces to a line element, which is conformal to the static metric $ds^2=\sqrt{\frac{F(r)}{G(r)}}ds^2_{static}$, where
\begin{equation}
    ds^2_{static}=-G(r)dt^2+\frac{dr^2}{F(r)}+H(r)d\Omega^2_{(2)},
\end{equation}
with $d\Omega^2_{(2)}$ is the two-dimensional unit sphere.

We consider  the special case $G(r)=F(r)\equiv g(r)$, {\color{black} according to} the static metric (\ref{e3}), and $H(r)=r^2$. We find that the metric (\ref{AA metric}) reduces to the following rotating line element

\begin{eqnarray}
ds^2 &=& -(\frac{g(r)r^2+a^2\cos^2{\theta}}{r^2+a^2\cos^2{\theta}})dt^2+(\frac{r^2+a^2\cos^2{\theta}}{g(r)r^2+a^2})dr^2+(r^2+a^2\cos^2{\theta})d\theta^2 \nonumber\\
     &+& (r^2+a^2\cos^2{\theta})\sin^2{\theta} \Bigl(1+a^2\sin^2{\theta}[\frac{2r^2-g(r)r^2+a^2\cos^2{\theta}}{(r^2+a^2\cos^2{\theta})^2}] \Bigl) d\phi^2
     \nonumber\\
     &-& 2a\sin^2{\theta}\Bigl( \frac{r^2-g(r)r^2}{r^2+a^2\cos^2{\theta}}\Bigl) dtd\phi. \label{erot}
\end{eqnarray}
 
We also note that In the limit $a=0$, this metric (\ref{erot}) reduces to the static spherically symmetric line element (\ref{e3}). We require the inherently rotating line element (\ref{erot}) to be a solution of $f(R)$ theory with the $f(R)$ function (\ref{fr}) and the metric function (\ref{e10}). We then find that the components of $T_{\mu\nu}^{(matter)}$ are non-zero. 
The only non-zero components of the energy-momentum tensor $T_{\mu\nu}^{(matter)}$ are $T_{tt}$, $T_{rr}$, $T_{\theta\theta}$, $T_{\phi\phi}$, $T_{r\theta}$ and $T_{t\phi}$,. We list all the non-zero components of the energy-momentum tensor in appendix A. 
We realize that $T_{\mu\nu}^{(matter)}$ has the form of an imperfect fluid 
\begin{equation}
    T_{\mu\nu}=\rho u_{\mu}u_{\nu}+q_{\mu}u_{\nu}+q_{\nu}u_{\mu}+\Pi_{\mu\nu},\label{TTm}
\end{equation}
with the effective fluid 4-velocity $u^{\mu}$, effective energy density $\rho=T_{\mu\nu}u^{\mu}u^{\nu}$, heat flux density $q_{\mu}$, and the stress tensor $\Pi_{\mu\nu}$. The stress tensor can be written in terms of the isotropic pressure $P$ and anisotropic stresses $\pi_{\mu\nu}$ as 
\begin{equation}
    \Pi_{\mu\nu}=Ph_{\mu\nu}+\pi_{\mu\nu},
\end{equation}
where we introduced the projection operator $h_{\mu} \ ^{\nu}$ to split the spacetime into the time direction $u^{\mu}$ and the 3-dimensional space of the co-moving observer with the fluid $h_{\mu\nu}\equiv g_{\mu\nu}+u_{\mu}u_{\nu}$ \cite{ellis2009republication}.

\textcolor{black}{In order to evaluate the energy conditions in $f(R)$ gravity, We rewrite the field equation (\ref{e2}) in the form of the effective Einstein field equation}
\begin{equation}
    R_{\mu\nu}=\mathcal{T}_{\mu\nu}-\frac{1}{2}g_{\mu\nu}\mathcal{T},
\end{equation}
\textcolor{black}{where {\color{black} $\mathcal{T}_{\mu\nu}$ is the effective energy-momentum tensor} and $\mathcal{T}=g^{\mu\nu}\mathcal{T}_{\mu\nu}$. The {\color{black} effective energy-momentum tensor} $\mathcal{T}_{\mu\nu}$ and $\mathcal{T}$ are given by}
\begin{equation}
     \mathcal{T}_{\mu\nu}=\frac{1}{f_R}(T^{(matter)}_{\mu\nu}+\nabla_{\mu}\nabla_{\nu}f_R-g_{\mu\nu}\Box f_R),
\end{equation}
\begin{equation}
    \mathcal{T}=\frac{1}{f_R}(T^{(matter)}+f(R)-Rf_R-3\Box f_R).
\end{equation}
\textcolor{black}{In these equations $T^{(matter)}=g^{\mu\nu}T^{(matter)}_{\mu\nu}$ and $T^{(matter)}_{\mu\nu}$ is given in (\ref{TTm}), and we show its components explicitly in the appendix A.} In the observer comoving frame, the energy conditions are given by \cite{pimentel2016energy, maeda2023energy, NEW}
 \begin{equation}
     \epsilon_1\equiv \rho_{(eff)} \geq 0,\label{ee1}
 \end{equation}
\begin{equation}\label{ee2}
   \epsilon_2\equiv \rho_{(eff)}+3\hat{p}_{(eff)} \geq0,
\end{equation}
\begin{equation}\label{ee3}
    \epsilon_3\equiv\rho_{(eff)}-q_{(eff)} \geq 0,
\end{equation}
which represent weak, strong and dominant energy conditions, respectively. \textcolor{black}{Note that these quantities $\rho_{(eff)}$, $\hat{p}_{(eff)}$ and $q_{(eff)}$ corresponds to the imperfect fluid form of the effective energy-momentum tensor $\mathcal{T}_{\mu\nu}$.} In equations (\ref{ee2}) and (\ref{ee3}), $q_{(eff)}=\sqrt{q_{1(eff)}^2+q_{2(eff)}^2+q_{3(eff)}^2}$ and in the frame co-moving with the fluid $\hat{p}_{(eff)}=\frac{p^{(eff)}_1+p^{(eff)}_2+p^{(eff)}_3}{3}$.
We numerically verify that for a large range of variables, the energy-momentum tensor for this black hole solution satisfies all the mentioned energy conditions. 
In figure \ref{WEC}, we plot the effective energy density $\rho_{(eff)}$ versus the coordinates $r$ and $\theta$ for different values of the rotational parameter, cosmological constant and the mass parameter. We notice in all three cases, the  energy density $\rho_{(eff)}$ is positive definite and reaches to its minimum of zero at $\theta=\frac{\pi}{2}$. Hence the weak energy condition (\ref{ee1}) is satisfied for the effective energy-momentum tensor equation $\mathcal{T}_{\mu\nu}$.

\begin{figure}[h]\label{rho}
	\centering
	\begin{subfigure}{0.32\linewidth}
		\includegraphics[width=\linewidth]{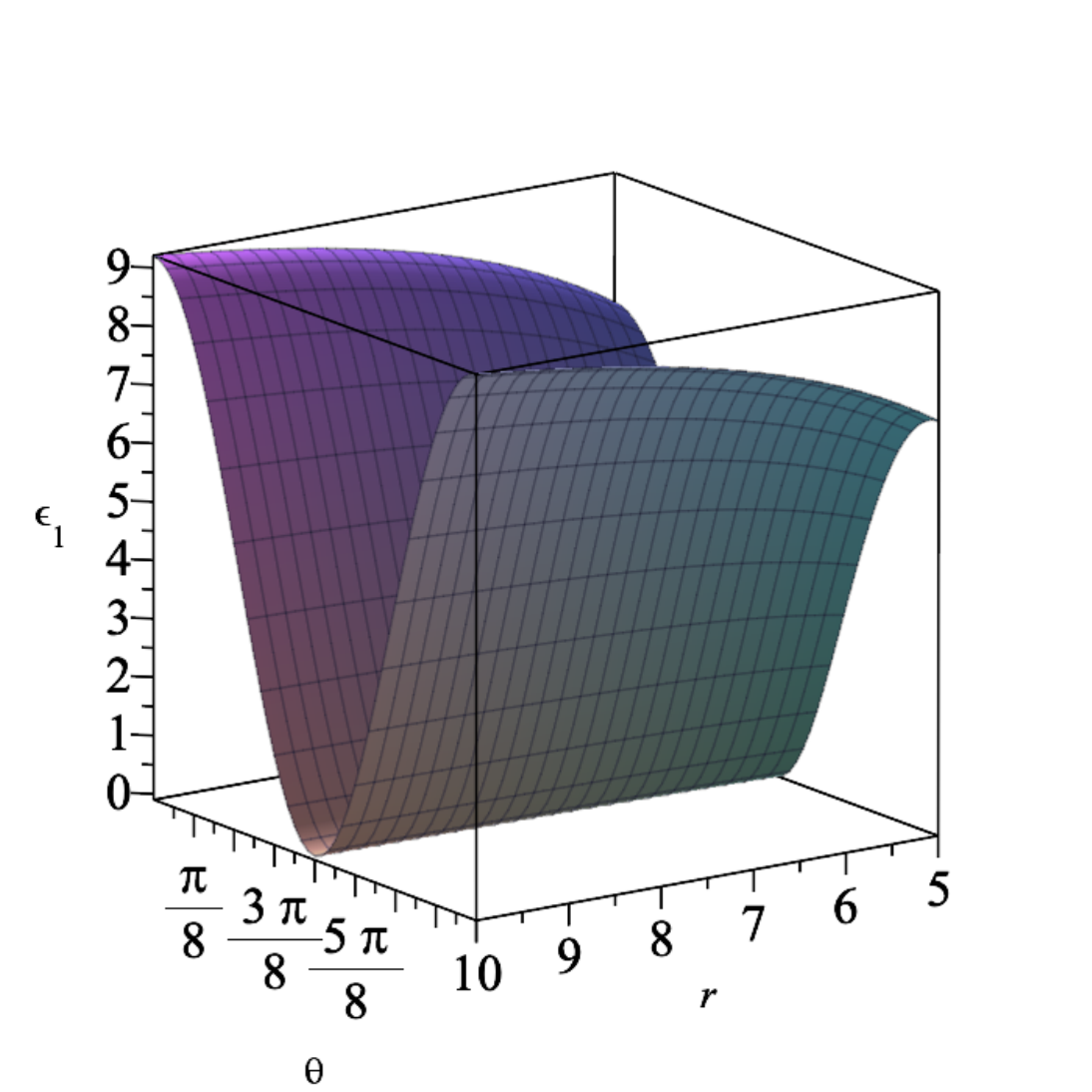}
		\caption{\centering}
		\label{}
	\end{subfigure}
       \begin{subfigure}{0.32\linewidth}
		\includegraphics[width=\linewidth]{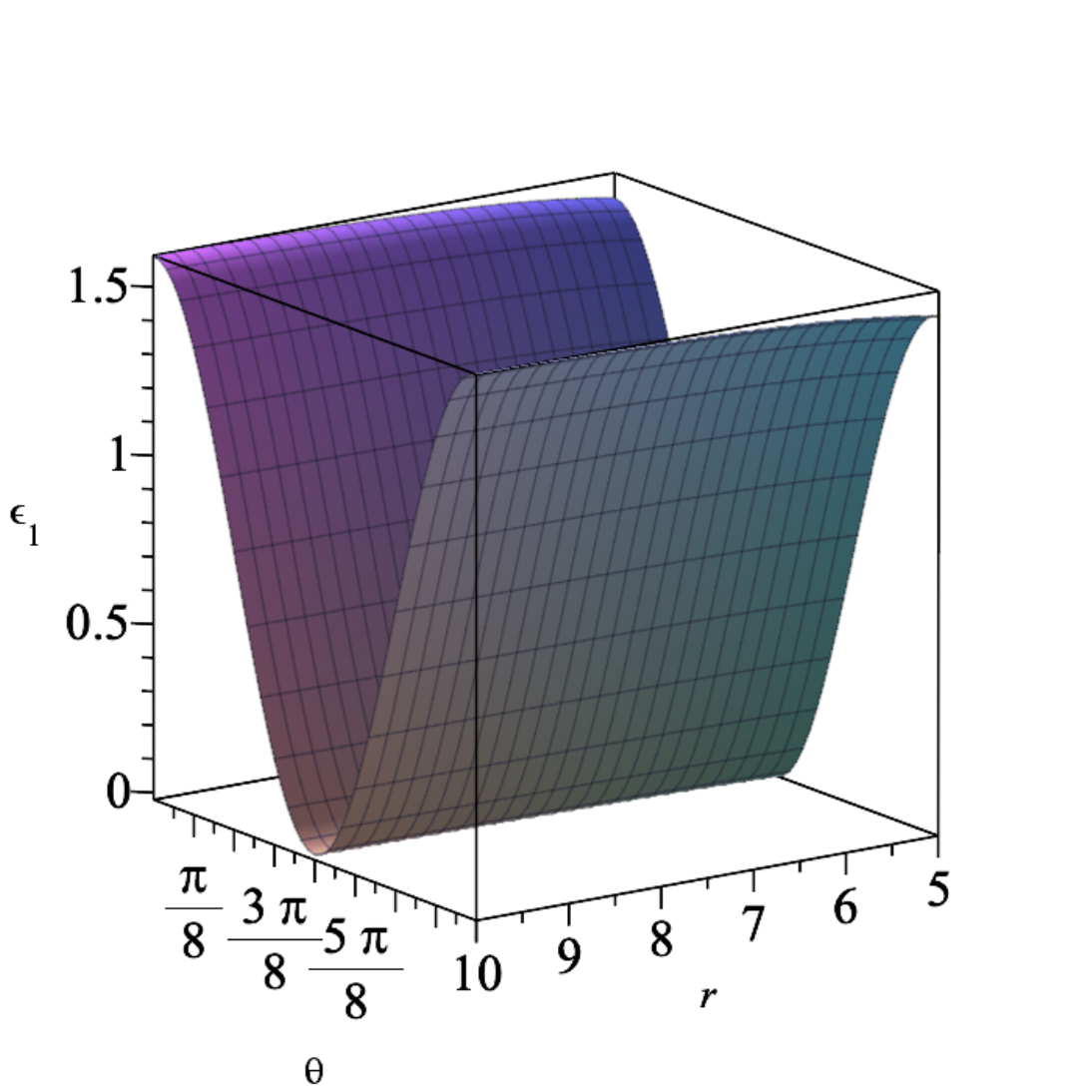}
		\caption{\centering}
		\label{}
	\end{subfigure}
        \begin{subfigure}{0.32\linewidth}
		\includegraphics[width=\linewidth]{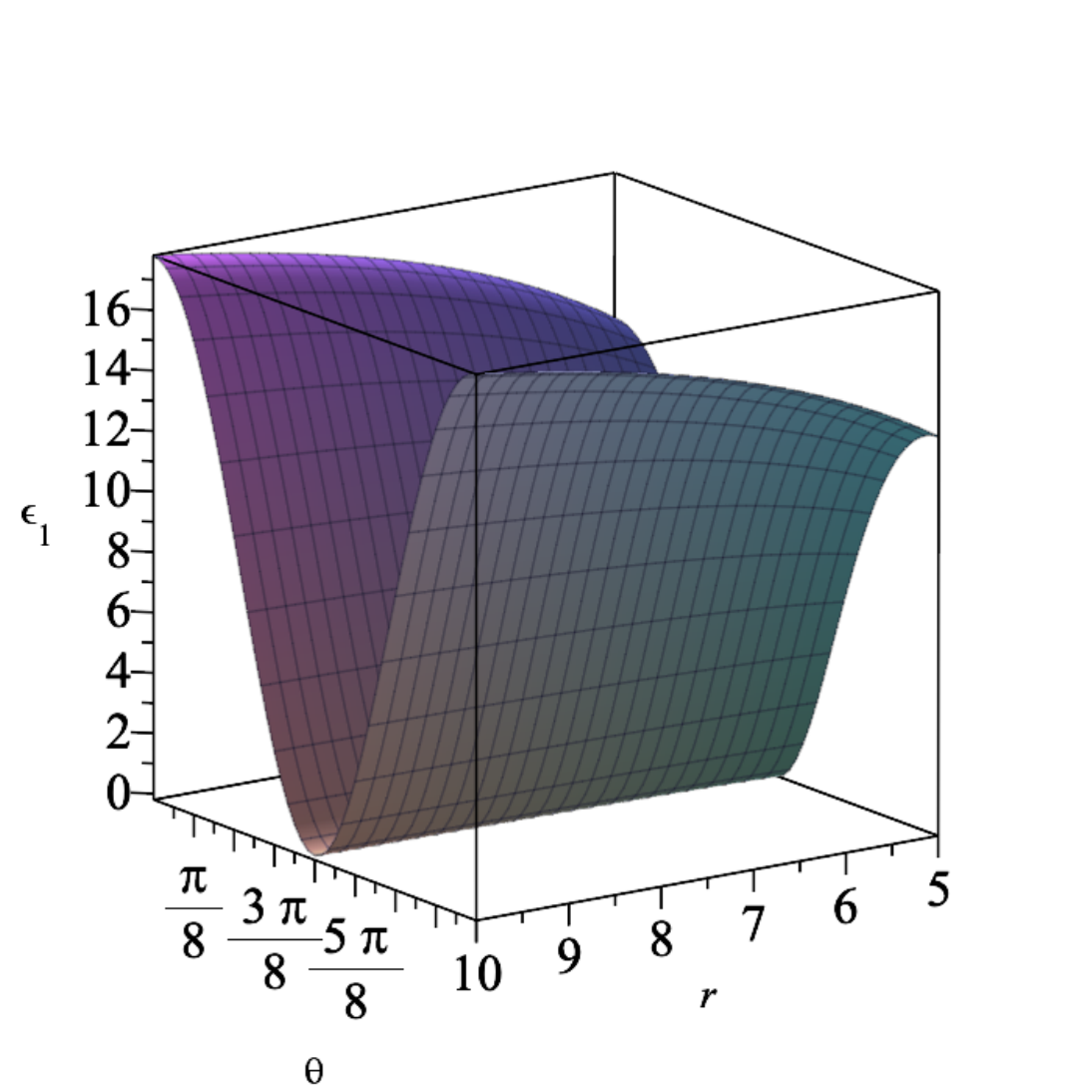}
		\caption{\centering} 
		\label{}
	\end{subfigure}
	\caption{Weak energy condition: The behaviour of $\rho_{(eff)}$ with respect to the coordinates $r$ and $\theta$, where (a): $a=2$, $\Lambda=-1.5$ and $M=10$ (b): $a=1$, $\Lambda=-1.2$ and $M=5$ (c): $a=2.5$, $\Lambda=-1.7$ and $M=1$. }
	\label{WEC}
\end{figure}

In figure \ref{SEC}, we plot the left hand side of the strong energy condition, i.e.  $\rho_{(eff)}+3\hat p_{(eff)}$ versus the coordinates $r$ and $\theta$ for different values of the rotational parameter, cosmological constant and the mass parameter. We notice in all three cases, the strong energy condition holds and it reaches to its minimum of zero at $\theta=\frac{\pi}{2}$. Hence the strong energy condition (\ref{ee2}) is satisfied for the energy-momentum tensor equation (\ref{TTm}).

\begin{figure}[h]
	\centering
	\begin{subfigure}{0.32\linewidth}
		\includegraphics[width=\linewidth]{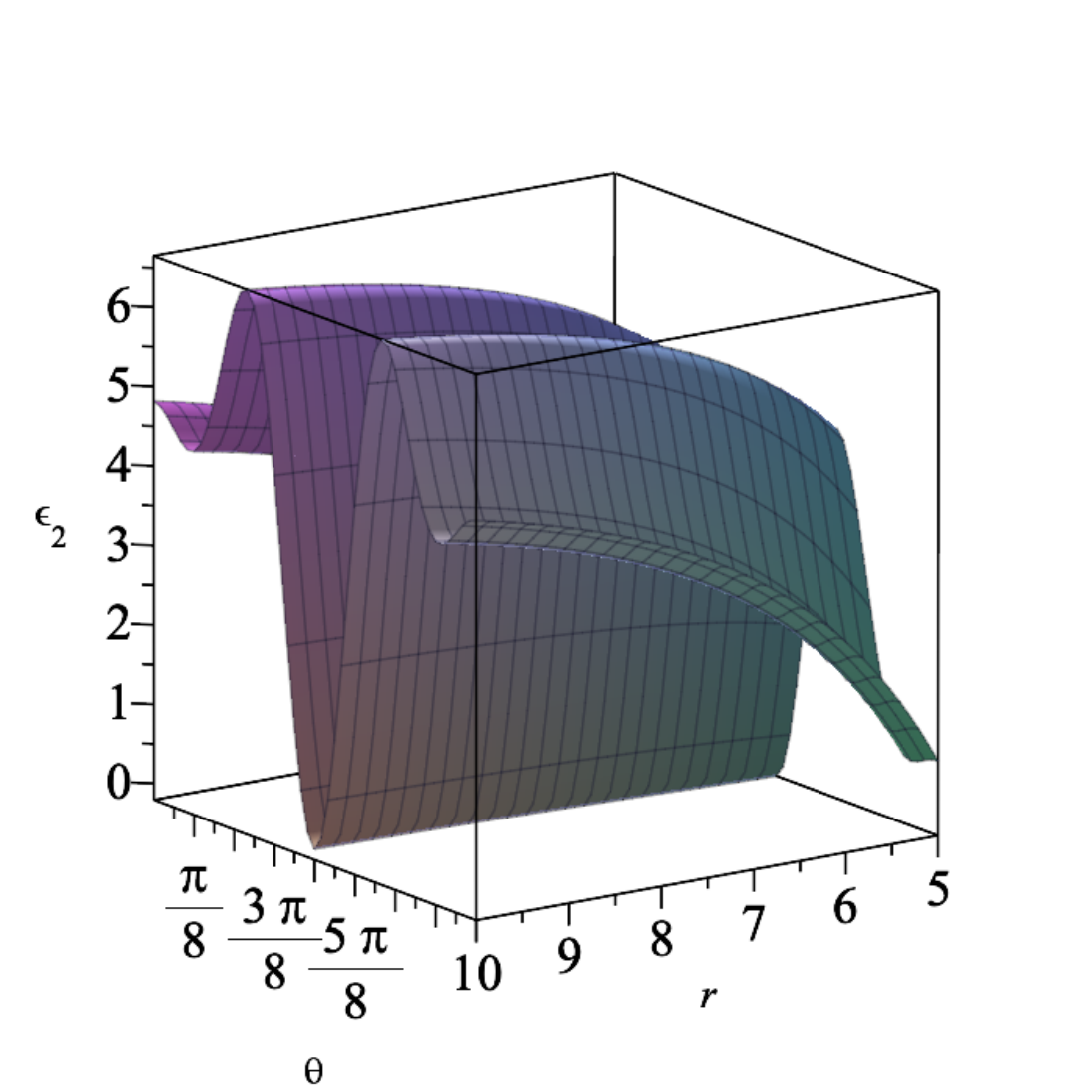}
		\caption{\centering}
		\label{}
	\end{subfigure}
       \begin{subfigure}{0.32\linewidth}
		\includegraphics[width=\linewidth]{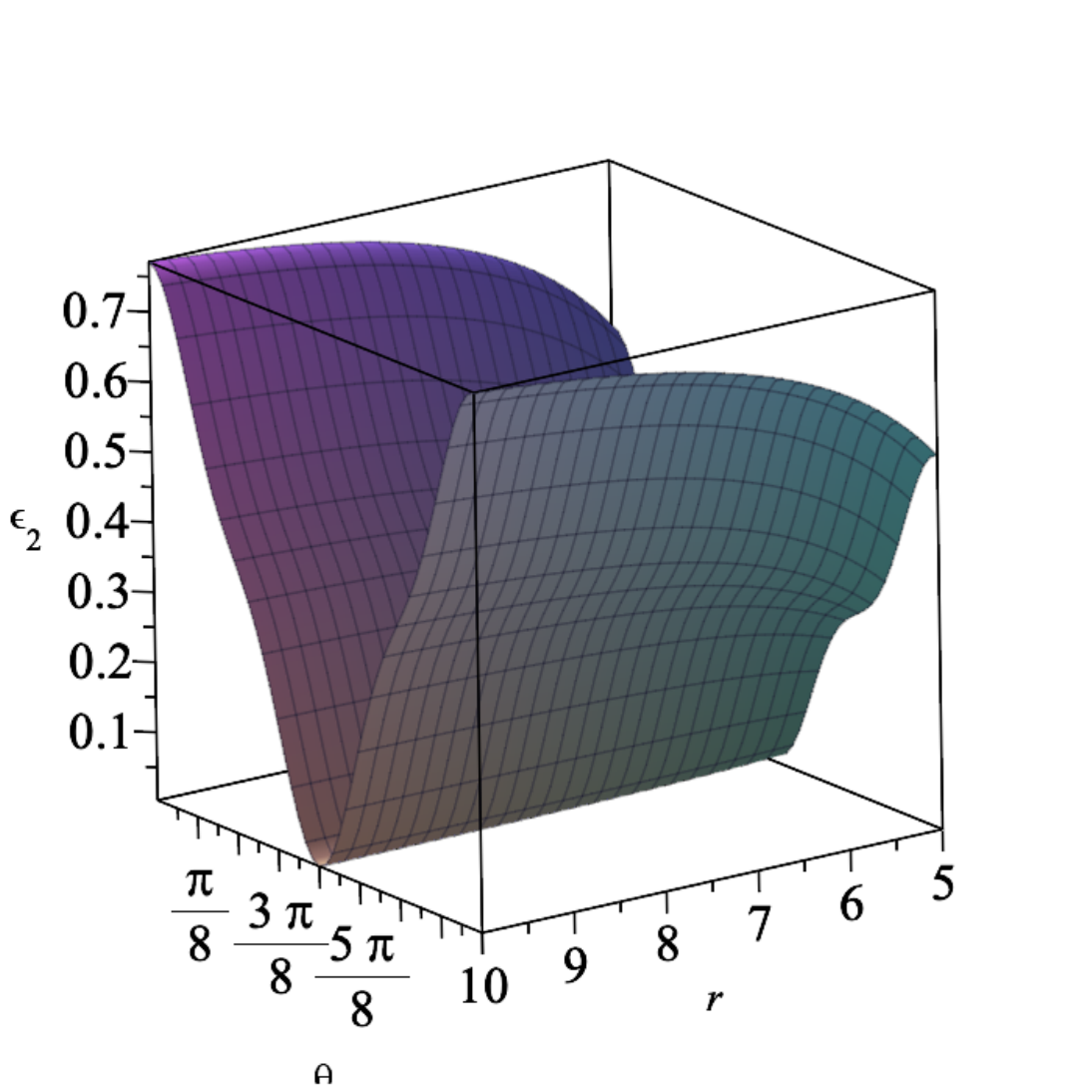}
		\caption{\centering}
		\label{}
	\end{subfigure}
        \begin{subfigure}{0.32\linewidth}
		\includegraphics[width=\linewidth]{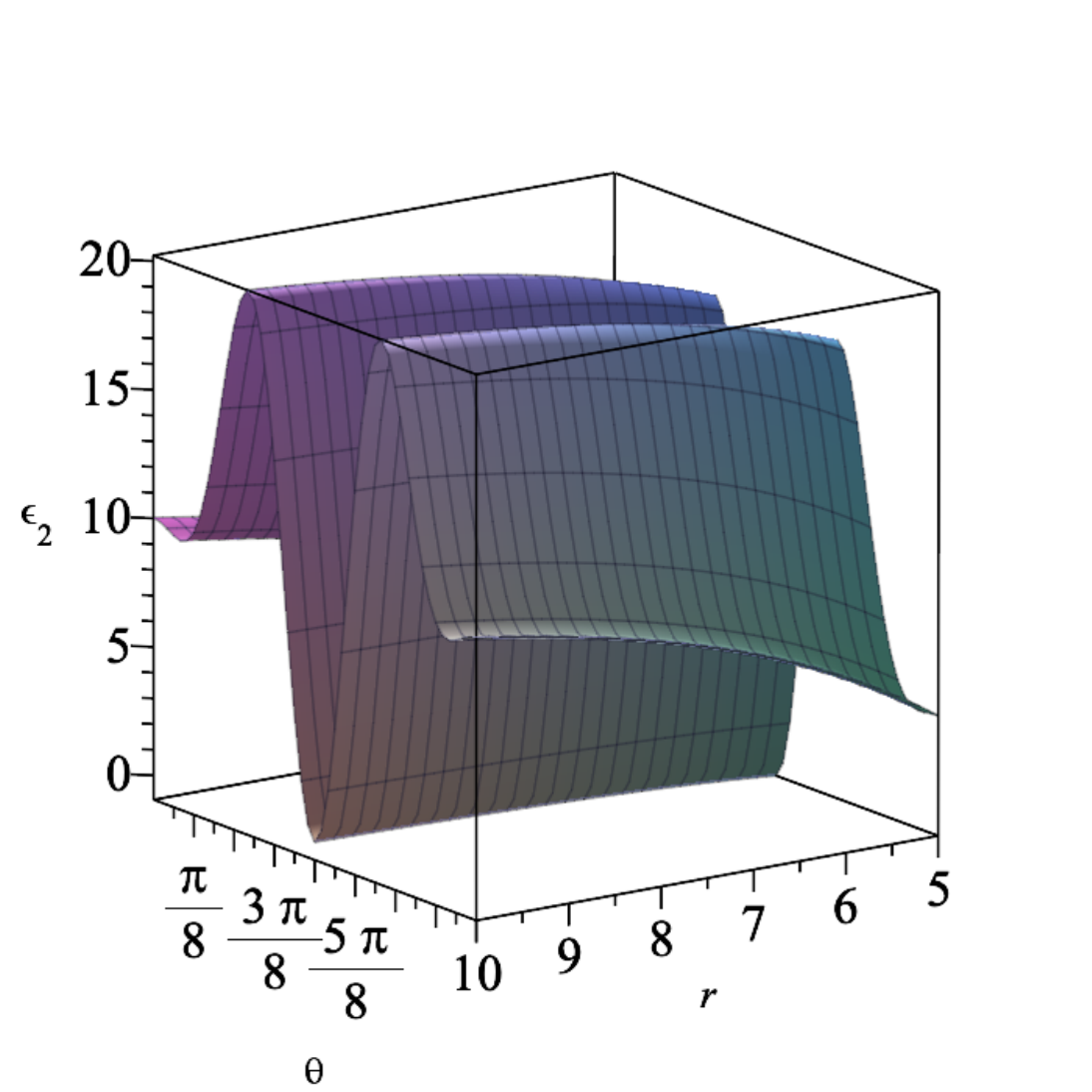}
		\caption{\centering} 
		\label{}
	\end{subfigure}
	\caption{Strong energy condition: The behaviour of $\rho_{(eff)}+3\hat{p}_{(eff)}$ with respect to the coordinates $r$ and $\theta$, where (a): $a=2$, $\Lambda=-1.5$ and $M=10$ (b): $a=1$, $\Lambda=-1.2$ and $M=5$; (c): $a=2.5$, $\Lambda=-1.7$ and $M=1$. }
	\label{SEC}
\end{figure}

In figure \ref{DEC}, we plot the left hand side of the dominant energy condition, i.e.  $\rho_{(eff)}-q_{(eff)}$ versus the coordinates $r$ and $\theta$ for different values of the rotational parameter, cosmological constant and the mass parameter. We notice in all three cases, the dominant energy condition holds and it reaches to its minimum of zero at $\theta=\frac{\pi}{2}$. Hence the dominant energy condition (\ref{ee3}) is satisfied for the energy-momentum tensor equation (\ref{TTm}).

\begin{figure}[h]
	\centering
	\begin{subfigure}{0.32\linewidth}
		\includegraphics[width=\linewidth]{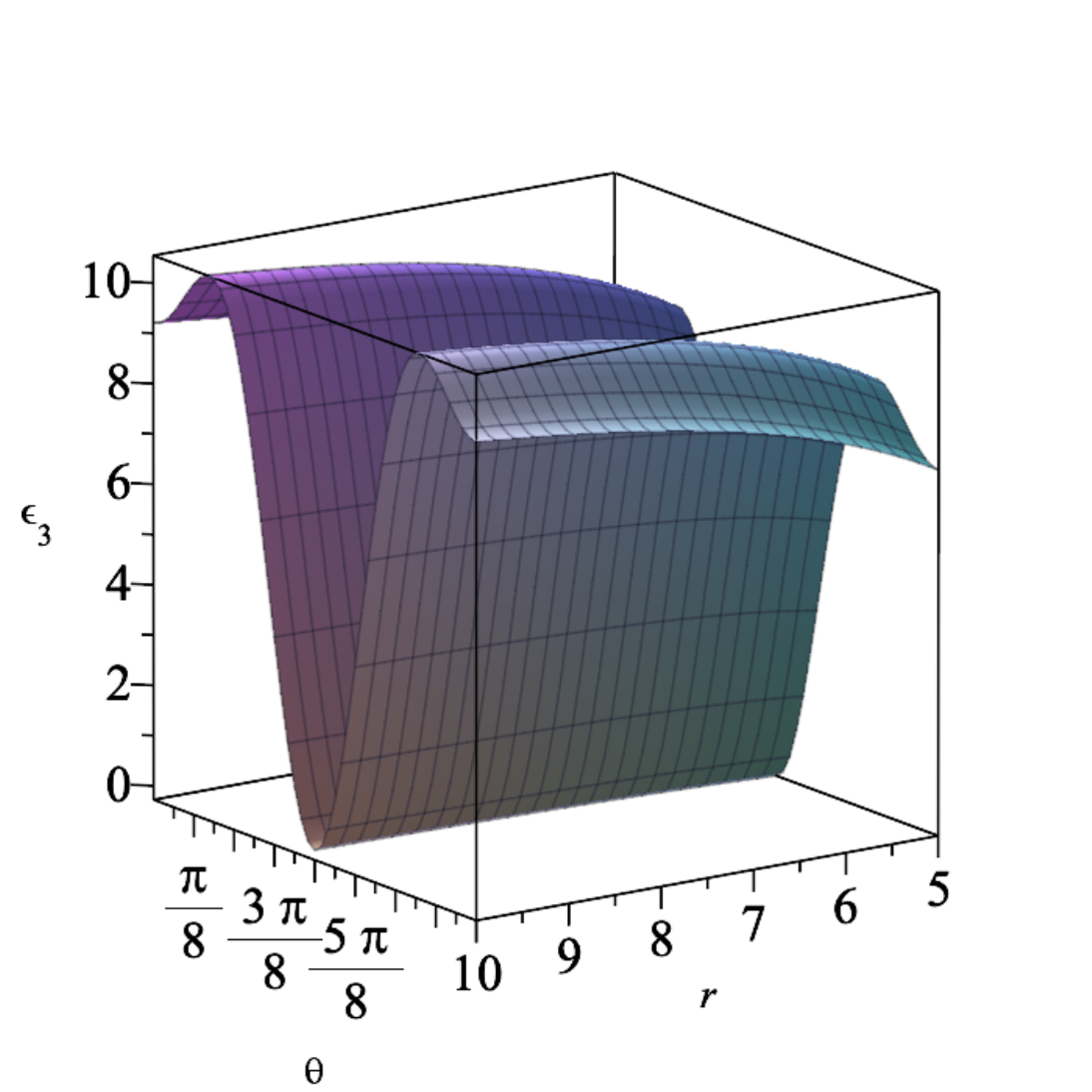}
		\caption{\centering}
		\label{}
	\end{subfigure}
       \begin{subfigure}{0.32\linewidth}
		\includegraphics[width=\linewidth]{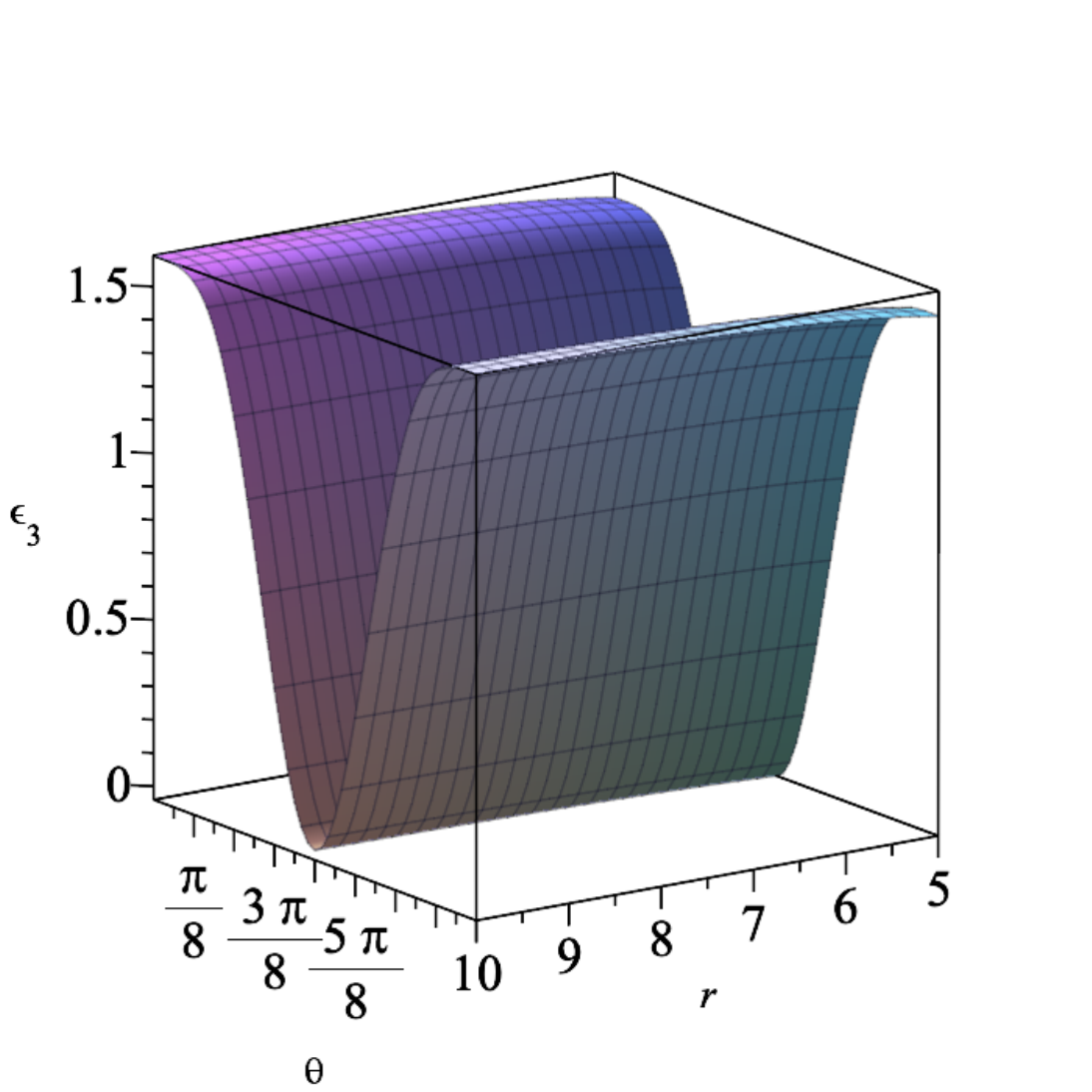}
		\caption{\centering}
		\label{}
	\end{subfigure}
        \begin{subfigure}{0.32\linewidth}
		\includegraphics[width=\linewidth]{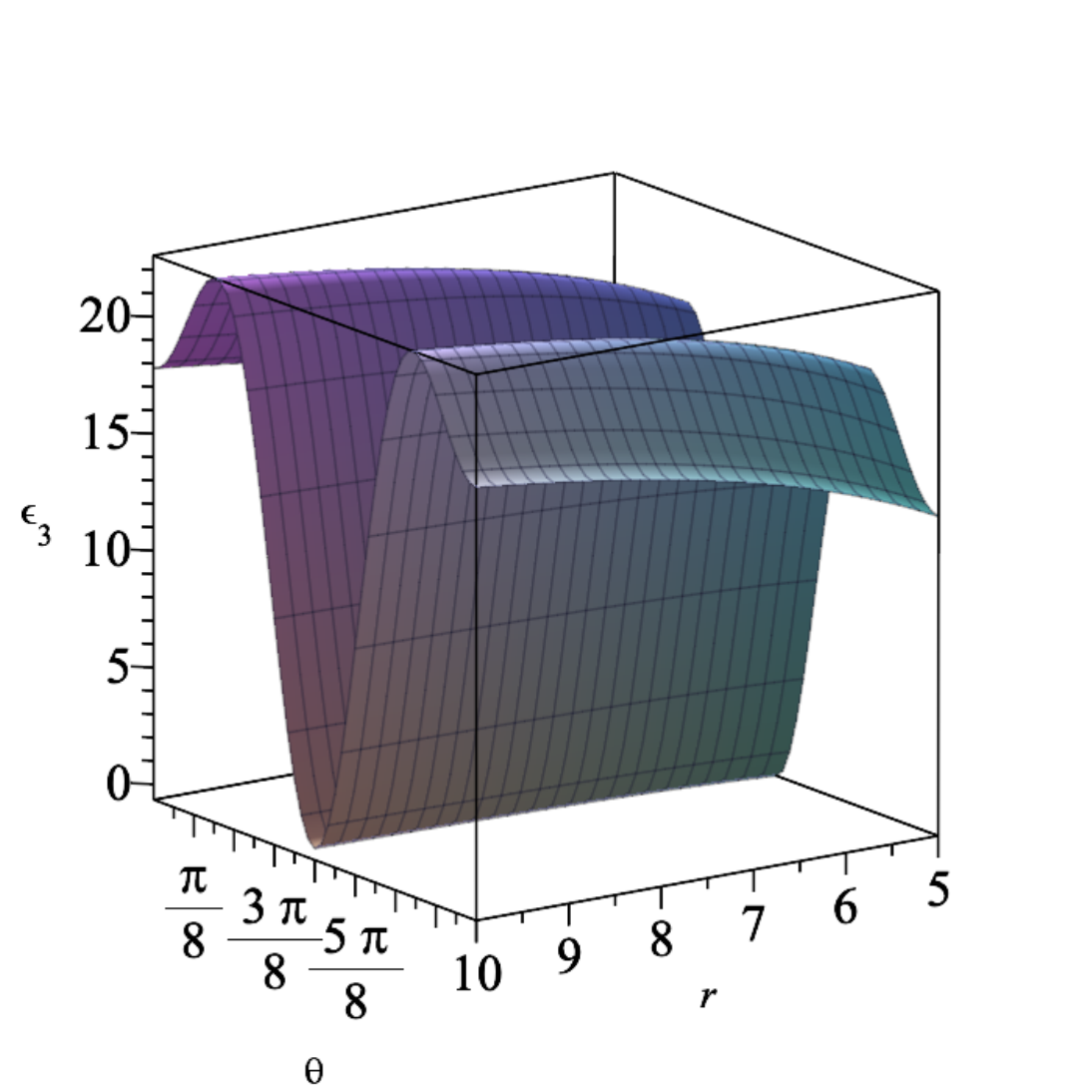}
		\caption{\centering} 
		\label{}
	\end{subfigure}
	\caption{Dominant energy condition: The behaviour of $ \rho-q$ with respect to the coordinates $r$ and $\theta$, where (a): $a=2$, $\Lambda=-1.5$ and $M=10$ (b): $a=1$, $\Lambda=-1.2$ and $M=5$ (c): $a=2.5$, $\Lambda=-1.7$ and $M=1$. }
	\label{DEC}
\end{figure}

\subsection{Hidden conformal symmetry for the type I rotating black hole solution in $f(R)$ gravity}

In order to find the hidden conformal symmetry dual to the type I rotating black hole solution (\ref{erot}), we consider a massless scalar probe $\Phi$ in the background of the black hole (\ref{erot}). We try to find the hidden conformal symmetry in the solution space of the probe field in the background of a generic black hole (\ref{erot}). We also should note that the near horizon geometry of the generic black hole (\ref{erot}) does not possess any AdS isometries \cite{castro2010hidden}. The Klein-Gordon equation for the massless scalar field $\Phi$ is given by
\begin{equation}
    \frac{1}{\sqrt{-g}}\partial_{\mu}(\sqrt{-g}g^{\mu\nu}\partial_{\nu}\Phi)=0, \label{e13}
\end{equation}
where $g$ is the determinant of the metric $g_{\mu\nu}$. Since the line element (\ref{erot}) has two Killing vectors, we separate the coordinates in the scalar field as
\begin{equation}
    \Phi=\exp{(-i\omega t+im\phi)}R(r)S(\theta), \label{e14}
\end{equation}
where $R(r)$ and $S(\theta)$ are the radial and angular function of the probe, respectively. Plugging equation (\ref{e14}) into the Klein-Gordon equation (\ref{e13}) leads to two differential equations for the radial $R(r)$ and angular $S(\theta)$ wave functions
\begin{equation}
    (g(r)r^2+a^2)\frac{d^2R(r)}{dr^2}+(r^2\frac{dg(r)}{dr}+2rg(r))\frac{dR(r)}{dr}+V(r)R(r)=0, \label{e15}
\end{equation}
\begin{equation}
    \frac{1}{S(\theta)}\frac{d^2S(\theta)}{d\theta^2}+\frac{\cos{\theta}}{\sin{\theta}S(\theta)}\frac{dS(\theta)}{d\theta}+\frac{a^2\omega^2\cos^4{\theta}+a^2\omega^2-m^2}{\sin^2{\theta}}=0,
\end{equation}
where
\begin{equation}
    V(r)=-\frac{2g(r)a^2\omega^2r^2-2g(r)am\omega r^2+a^4\omega^2-2a^2\omega^2r^2-\omega^2r^4+2ar^2m\omega-a^2m^2}{g(r)r^2+a^2}.
\end{equation}

The positive roots of the metric function $\Delta(r)\equiv g(r)r^2+a^2$ indicates the location of the horizons of the black hole. In order to establish the hidden conformal symmetry, we {\color{black} compare} the radial wave equation (\ref{e15}) of the probe to the Casimir operator of a CFT 
\cite{sakti2020hidden, chen2010holographic}. Therefore, it is necessary to expand the metric function $\Delta(r)$ in the near-horizon region as a quadratic polynomial in $(r-r_+)$, where $r_+$ is the outer horizon of the black hole. We expand the metric function $\Delta(r)$ such as
\begin{equation}
    \Delta(r)\simeq K(r-r_+)(r-r_*), \label{expan}
\end{equation}
 where $K$ and $r_*$ are constants, that depend on the form of the metric function $\Delta(r)$ and the black hole parameters, and are given by
\begin{equation}
    K=1-2\Lambda r_+^2,
\end{equation}
\begin{equation}
    r_*=r_++\frac{2r_+-M-4/3\Lambda r_+^3}{2\Lambda r_+^2-1}.
\end{equation}
We note that  $r_*$ is not necessarily a horizon of the black hole.

We consider the radial equation (\ref{e15}) at the near-horizon region, where $\omega r_+ \ll 1$. Also, we consider a limit where the outer horizon $r_+$ is very close to ${r_ * }$. With these approximations, the radial equation (\ref{e15}) simplifies to
\begin{equation}
   \frac{d}{dr}[\left( {r - {r_ + }} \right)\left( {r - {r_ * }} \right)\frac{d}{dr}R\left( r \right)]+ \left[ {\left( {\frac{{{r_ + } - {r_ * }}}{{r - {r_ + }}}} \right)\mathcal{A} + \left( {\frac{{{r_ + } - {r_ * }}}{{r - {r_ * }}}} \right)\mathcal{B} + \mathcal{C}} \right]R\left( r \right) = 0, \label{e19}
\end{equation}
where the explicit form of the constants $\mathcal{A}$, $\mathcal{B}$ and $\mathcal{C}$ are given in the Appendix B. 
We also note that the exact solutions to the angular $S(\theta)$ wave function is given by
\begin{eqnarray}
S \left( \theta \right) &=&\{ s_1{\cal H}_C \left( 0,-{\frac{1}{2
}},\sqrt {-2\,{a}^{2}{\omega}^{2}+{m}^{2}},{\frac {{a}^{2}{\omega}^{2}
}{4}},-{\frac {{a}^{2}{\omega}^{2}}{4}}+{\frac {{m}^{2}}{4}}+{\frac{1}
{4}},{\frac {\cos \left( 2\,\theta \right) }{2}}+{\frac{1}{2}}
 \right)   \nonumber\\
 &+&
s_2{\cos \theta }{\cal H}_C
 \left( 0,{\frac{1}{2}},\sqrt {-2\,{a}^{2}{\omega}^{2}+{m}^{2}},{
\frac {{a}^{2}{\omega}^{2}}{4}},-{\frac {{a}^{2}{\omega}^{2}}{4}}+{
\frac {{m}^{2}}{4}}+{\frac{1}{4}},{\frac {\cos \left( 2\,\theta
 \right) }{2}}+{\frac{1}{2}} \right) \}  \nonumber\\
&\times& \left(\cos^2\theta - 1\right) ^{{\frac {1}{2}\sqrt {-2\,{a}^{2}{\omega}^{2}+{m}^{2}}}},
\end{eqnarray}
where ${\cal H}_C$ is the Heun function, and $s_1$ and $s_2$ are two arbitrary constants. In figure \ref{SS}, we plot the typical behaviour of the angular wave function versus $\theta$.
\begin{figure}[h]
	\centering
	\begin{subfigure}{0.35\linewidth}
		\includegraphics[width=\linewidth]{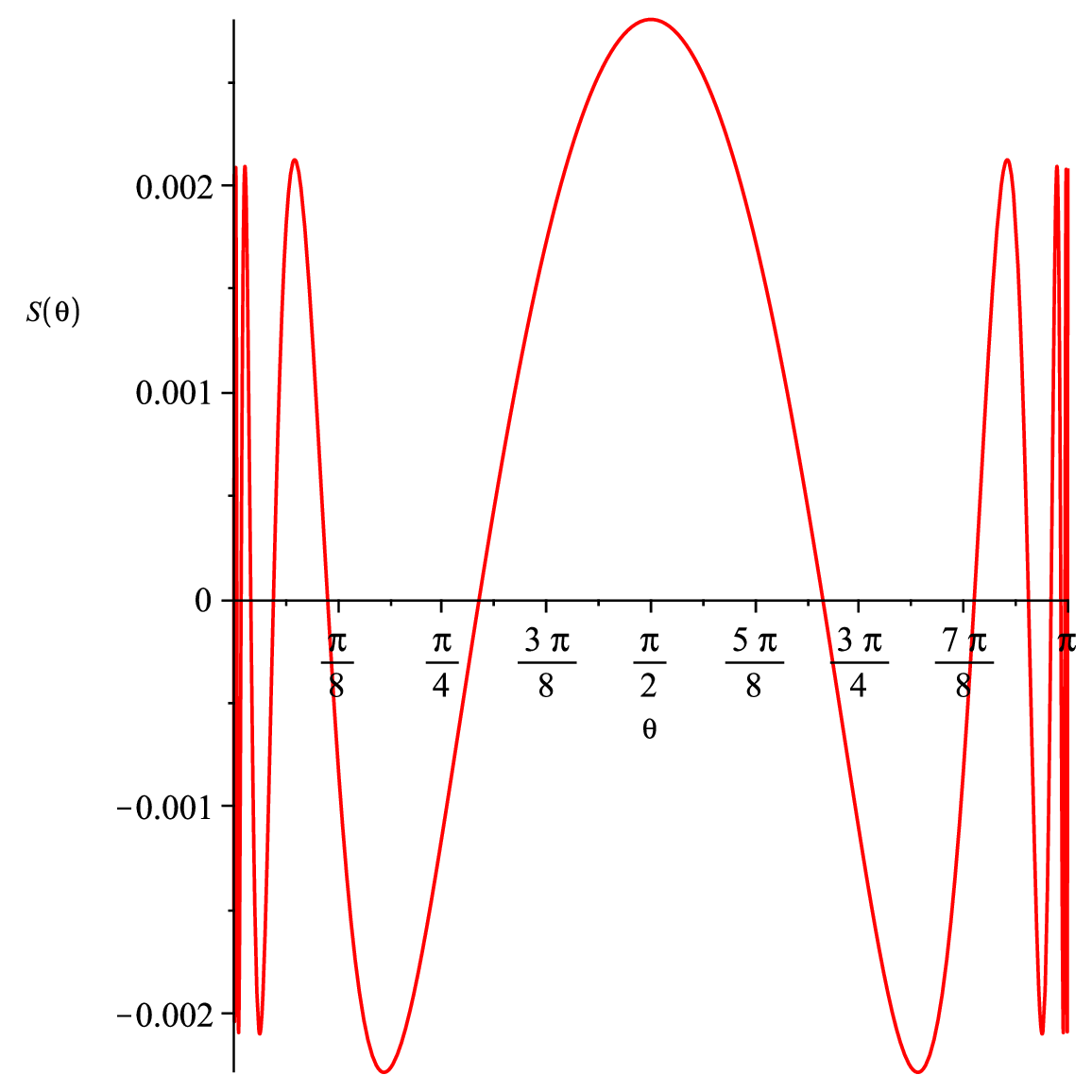}
		\caption{\centering}
		\label{}
	\end{subfigure}
       \begin{subfigure}{0.35\linewidth}
		\includegraphics[width=\linewidth]{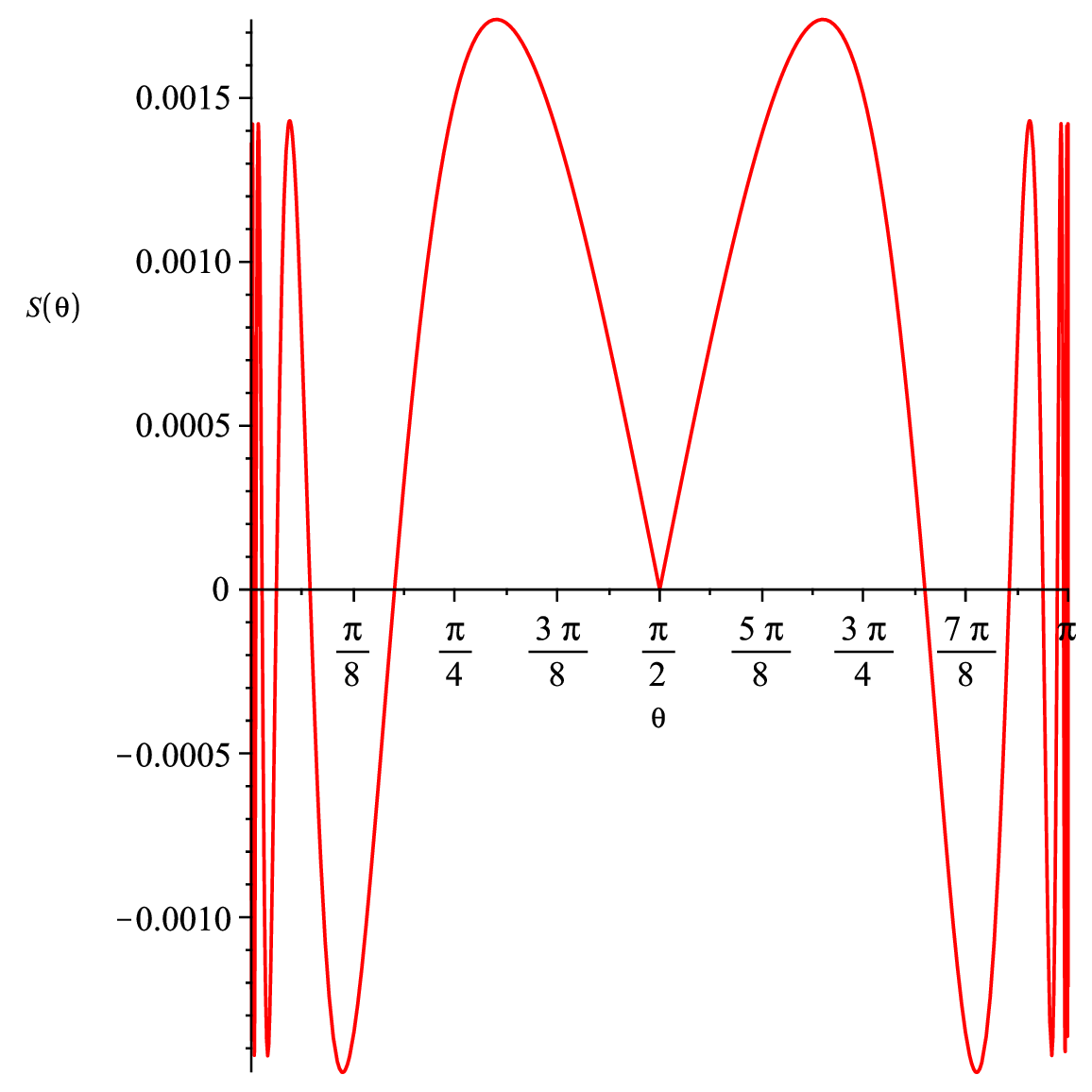}
		\caption{\centering}
		\label{}
	\end{subfigure}
	\caption{The behaviour of the angular wave function $S(\theta)$ versus $\theta$, for (a): $s_1=1,s_2=0$\, (b): $s_1=0,s_2=1$ where we set $a=1$, $m=2$, and $\omega=3$. }
	\label{SS}
\end{figure}

To investigate the existence of the hidden conformal symmetry, we consider the following conformal coordinates $\omega^+, \omega^-$ and $y$
\begin{eqnarray}
 \omega^{+}&=&\sqrt{\frac{r-r_{+}}{r-r_{*}}}e^{2\pi T_{R}\phi +2n_{R}t},
  \label{intro14}\\
  \omega^{-}&=&\sqrt{\frac{r-r_{+}}{r-r_{*}}}e^{2\pi T_{L}\phi +2n_{L}t},
  \label{intro15}\\
  y&=&\sqrt{\frac{r_{+}-r_{*}}{r-r_{*}}}e^{\pi( T_{R}+T_{L})\phi +(n_{R}+n_{L})t},
\label{intro16}
 \end{eqnarray} 
in terms of the black hole coordinates $t$, $r$ and $\phi$, where $T_L$, $T_R$, $n_L$ and $n_R$ are some parameters of the CFT.
We also define locally conformal operators as
\begin{equation}
    H_1=i\partial_{+} \label{opi},\  H_{-1}=i((\omega^{+})^2\partial_{+}+\omega^{+}y\partial_{y}-y^2\partial_{-}), \ H_0=i(\omega^{+}\partial_{+}+\frac{1}{2}y \partial_{y}),
\end{equation}
and
\begin{equation}
    \overline{H}_1=i\partial_{-}, \ \overline{H}_{-1}=i((\omega^{-})^2\partial_{-}+\omega^{-}y\partial_{y}-y^2\partial_{+}) \  \overline{H}_0=i(\omega^{-}\partial_{-}+\frac{1}{2}y\partial_{y}). \label{opf}
\end{equation}
The operators (\ref{opi}) and (\ref{opf}) satisfy the $SL(2,\mathbb{R})_L \times SL(2,\mathbb{R})_R$ algebra, which is given by
\begin{equation}
 ~~[H_0,H_{\pm1}]=\mp i H_{\pm 1},~~~~~~~~[H_{-1},H_1]=-2iH_0,~~
\label{intro21}
\end{equation}
\begin{equation}
 ~~[\overline{H}_0,\overline{H}_{\pm1}]=\mp i \overline{H}_{\pm 1},~~~~~~~~[\overline{H}_{-1},\overline{H}_1]=-2i\overline{H}_0.~~
\end{equation}
We can find the quadratic Casimir operators of $SL(2,\mathbb{R})_L \times SL(2,\mathbb{R})_R$ algebra, in terms of the conformal coordinates, from any two sets of the operators, as
\begin{eqnarray}
\mathcal{H}^2&=& \mathcal{\overline{H}}^2= -H_{0}^2+\frac{1}{2}(H_1H_{-1}+H_{-1}H_{1})\nonumber \\
&=&\frac{1}{4}
(y^2\partial_{y}^2-y\partial_{y})+y^2\partial_{+}\partial_{-}.
\label{intro28}
\end{eqnarray}
{\color{black} The quadratic Casimir  operator is a special element of the center of the universal enveloping algebra of a Lie algebra. A very well known example of the quadratic Casimir operator in physics, is the squared angular momentum operator, which is the Casimir element of the three-dimensional rotation algebra, and commutes with all the generators of the rotation algebra. The quadratic Casimir operator commutes with all the generators of the Lie algebra.}
We notice that the Casimir operator in terms of the coordinates $(t,r,\phi)$, is given by
\begin{eqnarray}
 \mathcal{H}^2&=&(r-r_+)(r-r_*)\partial_r^2+(2r-r_+-r_*)\partial_r+\frac{r_+-r_*}{r-r_*}(\frac{n_L-n_R}{4\pi G}\partial_{\phi}-\frac{T_L-T_R}{4G}\partial_t )^2 \nonumber\\
     &-& \frac{r_+-r_*}{r-r_+}(\frac{n_L+n_R}{4\pi G}\partial_{\phi}-\frac{T_L+T_R}{4G}\partial_t )^2, \label{csm}
\end{eqnarray}
where $G=n_LT_R-n_RT_L$. We can find that the radial wave equation (\ref{e19}) is, in fact, an eigenfunction-eigenvalue equation for the Casimir operator, such as 
\begin{equation}\label{ee}
    \mathcal{H}^2R(r)={\mathcal{\overline{H}}}^2R(r)=-{\cal C}R(r).
\end{equation}
Comparing (\ref{e19}) and (\ref{ee}), we find the CFT mode numbers and temperatures, are given by 
\begin{equation}
    n_L=\frac{U+1}{4V(\frac{X}{Y}-W)}, \label{e37a}
\end{equation}
\begin{equation}
    n_R=\frac{U-1}{4V(\frac{X}{Y}-W)},
\end{equation}
\begin{equation}
    T_L= \frac{-\frac{X}{Y}-1}{4\pi Z(-\frac{X}{Y}+W)},
\end{equation}
\begin{equation}
   T_R=\frac{-\frac{X}{Y}+1}{4\pi Z(-\frac{X}{Y}+W)}. \label{e40a}
\end{equation}
In equations (\ref{e37a})-(\ref{e40a}), the functions $U$, $V$, $X$, $Y$, $W$ and $Z$ are 
\begin{equation}
    U=\sqrt{\frac{r_+^6(r_+^2+2r_+r_*+3r_*^2)K^2+3Ka^2r_+^3r_*+a^4}{(3r_*^2r_+^6+2r_*^3r_+^5+r_*^4r_+^4)K^2+3Ka^2r_+^3r_*+a^4}},
\end{equation}

\begin{eqnarray}
    V &=& \sqrt{-\frac{r_*}{(r_+-r_*)^2(Kr_+^3r_*+a^2)^3Kr_+}}\Bigl(2K^3a^2r_+^9r_*^3+3K^2a^4r_+^6r_*^2-2K^2a^4r_+^5r_*^3
    \nonumber \\
    &-& K^2a^4r_+^4r_*^4-6K^2a^2r_+^8r_*^2 -4K^2a^2r_+^7r_*^3-2K^2a^2r_+^6r_*^4-3K^2r_+^{10}r_*^2-2K^2r_+^9r_*^3
    \nonumber \\
    &-& K^2r_+^8r_*^4+3Ka^6r_+^3r_*-6Ka^4r_+^5r_*-3Kr_+^7r_*a^2+a^8-2a^6r_+^2-a^4r_+^4\Bigl)^{1/2},
\end{eqnarray}

\begin{eqnarray}
    X &=& \Bigl( a^8+a^6(3Kr_+^3r_*-2r_+^2)+a^4(3K^2r_+^6r_*^2-2K^2r_+^7r_*-6Kr_+^5r_*-K^2r_+^8-r_+^4)
    \nonumber \\
    &+& a^2(2K^3r_+^9r_*^3-6K^2r_+^8r_*^2-4K^2r_+^9r_*-3Kr_+^7r_*-2K^2r_+^{10})-K^2r_+^{10}(r_+^2
    \nonumber \\
    &+& 2r_+r_*+3r_*^2)\Bigl)^{1/2},
\end{eqnarray}

\begin{eqnarray}
    Y &=& \Bigl( a^8+a^6(3Kr_+^3r_*-2r_+^2)+a^4(3K^2r_+^6r_*^2-2K^2r_+^5r_*^3-K^2r_+^4r_*^4-6Kr_+^5r_*-r_+^4)
    \nonumber \\
    &+& a^2Kr_+^6r_*(-2Kr_*^3+2K^2r_+^3r_*^2-4Kr_+r_*^2-6Kr_+^2r_*-3r_+)-K^2r_+^8r_*^2(3r_+^2
    \nonumber \\
    &+& 2r_+r_*+r_*^2)\Bigl)^{1/2},
\end{eqnarray}

\begin{equation}
    W=\sqrt{\frac{K^2r_+^6(r_+^2+2r_+r_*+3r_*^2)+3Ka^2r_+^3r_*+a^4}{K^2(3r_*^2r_+^6+2r_*^3r_+^5+r_*^4r_+^4)+3Ka^2r_+^3r_*+a^4}},
\end{equation}

\begin{equation}
    Z=\sqrt{\frac{a^2r_*(3K^2r_+^6r_*^2+2K^2r_+^5r_*^3+K^2r_+^4r_*^4+3Ka^2r_+^3r_*+a^4)}{(r_+-r_*)^2(Kr_+^3r_*+a^2)^3Kr_+}}.
\end{equation}

We show the typical behaviour of the right and left CFT temperatures versus the rotational parameter $a$, for different values of the cosmological constant $\Lambda$ in figure \ref{fig:f1}.

\begin{figure}[h]
	\centering
	\begin{subfigure}{0.4\linewidth}
		\includegraphics[width=\linewidth]{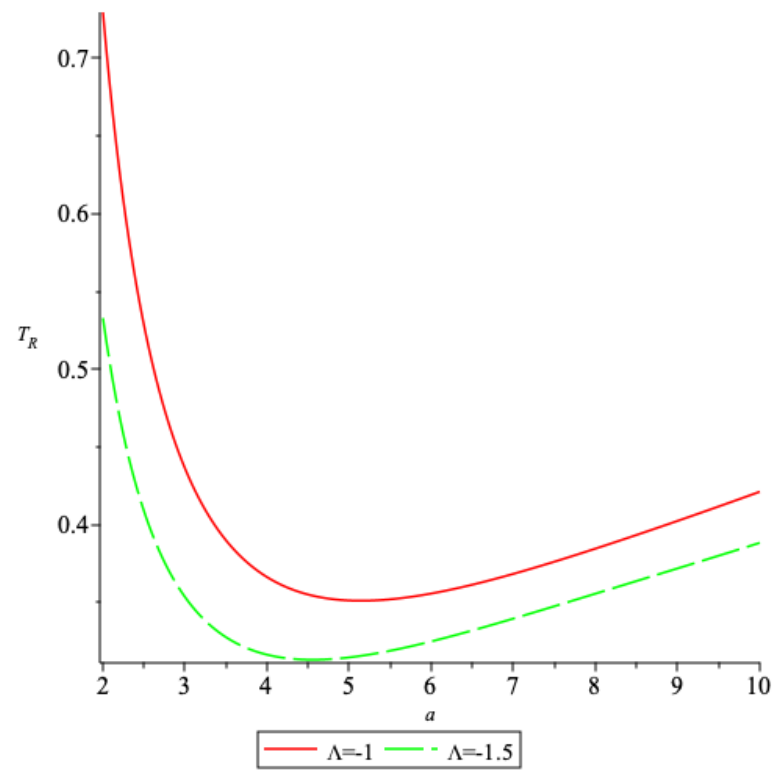}
		\caption{\centering}
		\label{fig:2a}
	\end{subfigure}
	\begin{subfigure}{0.4\linewidth}
		\includegraphics[width=\linewidth]{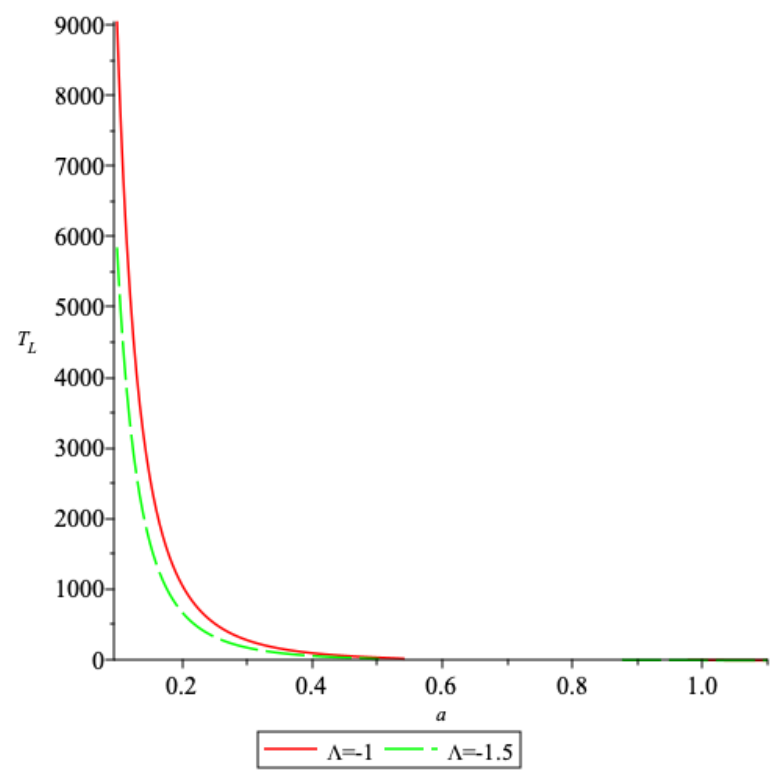}
		\caption{\centering} 
		\label{fig:2d}
	\end{subfigure}
	\caption{The behaviour of $T_R$ and $T_L$ for $M=10$ with respect to the rotation parameter. }
	\label{fig:f1}
\end{figure}

The $SL(2,\mathbb{R})_L \times SL(2,\mathbb{R})_R$ algebra generated by the vector fields (\ref{opi})-(\ref{opf}) is only a local hidden conformal symmetry and due to the spontaneous symmetry breakdown under the angular identification $\phi \sim \phi+2\pi$, it cannot be defined globally \cite{castro2010hidden}. Assuming that the near horizon dynamics can be described by a dual CFT correspondence, we find the left and right central charges of the  dual CFT by comparing the microscopic entropy of the dual CFT and the macroscopic Bekenstein-Hawking entropy of the black hole.

The entropy of the CFT  is given by the Cardy formula as
\begin{equation}
    S_{CFT}=\frac{\pi^2}{3}(c_LT_L+c_RT_R), \label{e41}
\end{equation}
where $c_L$ and $c_R$ are the  left and right central charges. 

It is known that in modified theories of gravity, the expression of the Bekenstein-Hawking entropy $S_{BH}$ is different than the one in Einstein general relativity. The $S_{BH}$ in $f(R)$ gravity in geometrized units, is given by \cite{N2}-\cite{zheng2018horizon}
\begin{equation}
    S_{BH}=\frac{A}{4}f'(R)|_{r_+}, \label{enth}
\end{equation}
where prime denotes derivative with respect to $R$, and $A$ is the area of event horizon, which is located at $g(r_+)r_+^2+a^2=0$. A lengthy calculation leads to the area of the event horizon, as given by   
\begin{equation}
    A=\int_0^{2\pi}\int_0^{\pi}\sqrt{-g} d\theta d\phi=4\pi(r_+^2+a^2),
\end{equation}
and the Bekenstein-Hawking entropy becomes
\begin{equation}
    S_{BH}=\pi(r_+^2+a^2)(1+\frac{16\Lambda^2}{3R^2}).
\end{equation}

We realize that for the following central charges, the Cardy entropy is in agreement with the Bekenstein-Hawking entropy
\begin{equation}
    c_{L}=c_R=\frac{2(r_+^2+a^2)(a^4\cos^4{\theta}+2a^2r_+^2\cos^2{\theta}+4r_+^4) Z(-\frac{X}{Y}+W)}{-\frac{X}{Y}r_+^4}.
\end{equation}

In figure \ref{fig:ent}, we show the behaviour of the entropy $S$ and central charges $c_{L,R}$ for different values of the rotational parameter $a$. As we notice, the entropy has its maximum values on the north and south pole. Moving from the north pole to the equator on horizon, the entropy decreases to its minimum on equator, and then increases to its maximum, on the south pole. The central charges are behaving opposite to the entropy . The central charges are minimum at the north pole and south pole on the horizon. Moving from the north pole to the equator on horizon, the central charges increase to their maxima on equator, and then decrease to their minima, on the south pole. 
\begin{figure}[h]
	\centering
	\begin{subfigure}{0.4\linewidth}
		\includegraphics[width=\linewidth]{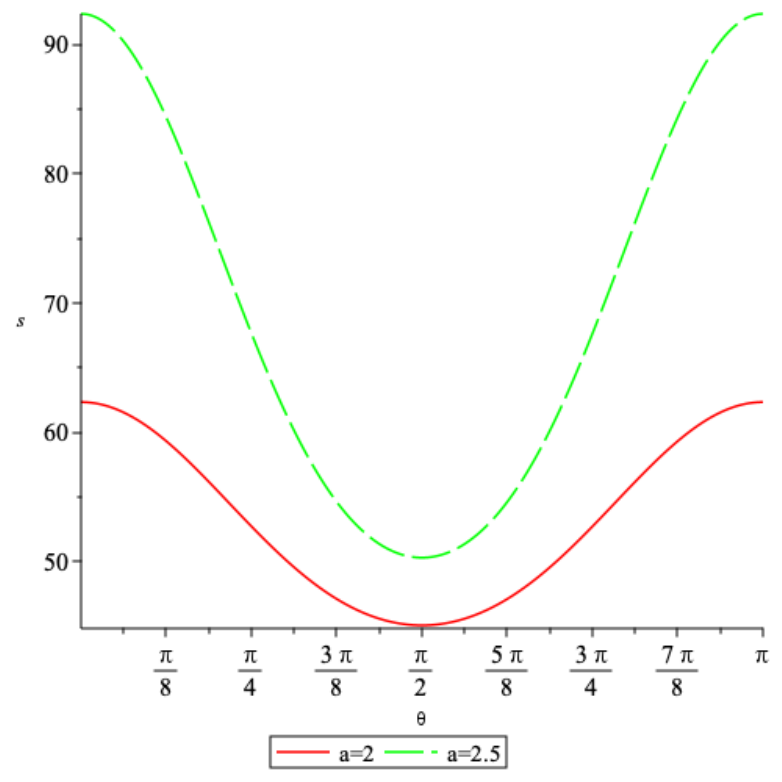}
		\caption{\centering}
		\label{fig:s}
	\end{subfigure}
	\begin{subfigure}{0.4\linewidth}
		\includegraphics[width=\linewidth]{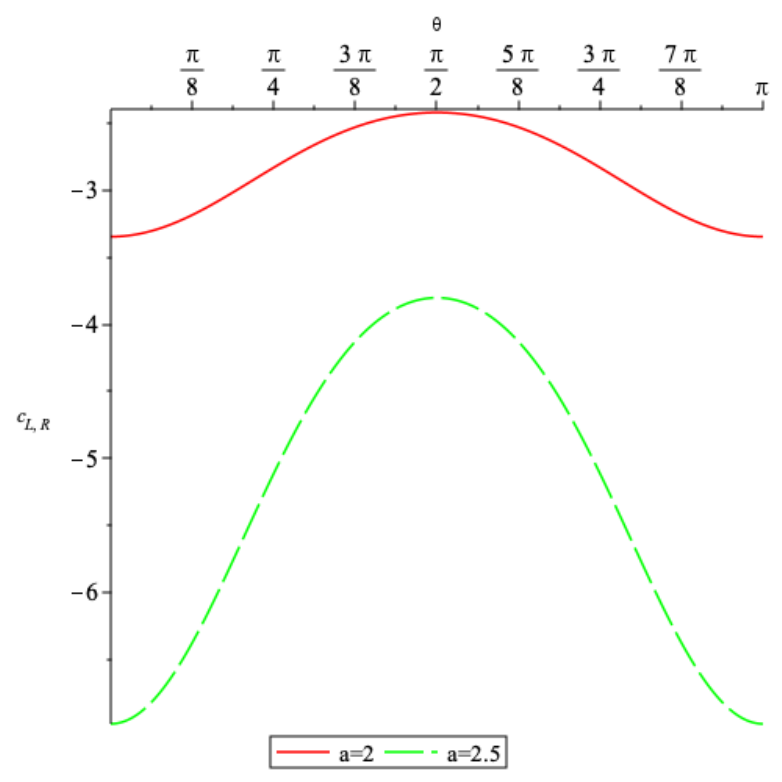}
		\caption{\centering} 
		\label{fig:c}
	\end{subfigure}
	\caption{The behaviour of the entropy $S$ and central charges $c_{L,R}$ for two different values of the rotational parameter $a$, where $M=10$ and $\Lambda=-1$. }
	\label{fig:ent}
\end{figure}
 

\section{TYPE II BLACK HOLE SOLUTIONS IN $f(R)$ GRAVITY}\label{sec:sec4}

\subsection{Boosted rotating black hole solutions in $f(R)$ gravity}

 Following the procedure developed and used in \cite{awad2003higher}-\cite{lemos1996cylindrical}, we apply the following Lorentz boost transformation
\begin{equation}
  \hat{\phi}=\Xi\phi+at,\label{e4a}
\end{equation}
\begin{equation}
\hat{t}=\Xi t+a\phi, \label{e4b}
\end{equation}
to the static spherically symmetric metric (\ref{e3}),  where $a$ is the local rotational parameter, and $\Xi=\sqrt{1+a^2}$. We find the boosted rotating line element as
\begin{eqnarray}
   ds^2&=& (-\Xi^2g(r)+a^2r^2\sin^2{\theta})dt^2+\frac{dr^2}{g(r)}+r^2d\theta^2+(\Xi^2r^2\sin^2{\theta}-a^2g(r))d\phi^2\nonumber\\
  &+ &  2a\Xi(r^2\sin^2{\theta}-g(r))dtd\phi,\label{e5}
\end{eqnarray}
where for simplicity, we dropped the overhead $\,\hat{}\,$ from the coordinates $\phi$ and $t$. The transformations (\ref{e4a}) and (\ref{e4b}) are not globally permitted, since they mix the compact and non-compact coordinates \cite{nashed2020new}. However, the transformations  (\ref{e4a}) and (\ref{e4b}) do not change the local properties of spacetime. Therefore, metrics (\ref{e3}) and (\ref{e5}) can be mapped locally into each other. Based on this reasoning, the transformations  (\ref{e4a}) and (\ref{e4b}) generate a new locally rotating metric \cite{lemos1996cylindrical, stachel1982globally}. We also note that similar transformations have been used to generate rotating solutions in general relativity and the MTG \cite{lemos1995three, awad2019rotating, kanzi2023superradiant}.

We explicitly verify that the boosted rotating line element (\ref{e5}) is a solution to the $f(R)$ field equations (\ref{e2}), where $f(R)$ is given by equation (\ref{fr})
\begin{equation}
    f(R)=R-\frac{16\Lambda^2}{3R},
\end{equation}
and the metric function $g(r)$ is given by equation (\ref{e10})
\begin{equation}
    g(r)=1-\frac{M}{r}-\frac{\Lambda}{3}r^2,
\end{equation}
and $T_{\mu\nu}^{(matter)}=0$. 
We also note that the Ricci scalar is a constant $R=4\Lambda$. Moreover, by calculating the Kretschmann invariant, we find that the curvature singularity appears only at $r=0$.
 The area $A$ of the black hole is given by
\begin{equation}
    A=\int_0^{2\pi}\int_0^{\pi}\sqrt{-g} d\theta d\phi=4\pi r_+^2\Xi,
\end{equation}
where $r_+$ is the outer horizon (positive root of $g(r)$). We calculate the Hawking temperature of the boosted black hole as (\ref{e5})
\begin{equation}
    T=\frac{1}{4\pi}|\frac{-2\Lambda}{3}r_++\frac{M}{r_+^2}|.
\end{equation}

\subsection{Hidden conformal symmetry for a rotating black hole solution in $f(R)$ gravity}
We consider a massless scalar probe in the background of the black hole (\ref{e5}). Due to the existence of two Killing vectors in the line element (\ref{e5}), we separate the coordinates in the scalar field $\Phi$ as
\begin{equation}
    \Phi=\exp{(-i\omega t+im\phi)}R(r)S(\theta),
\end{equation}
where $R(r)$ and $S(\theta)$ are the radial and angular function of the scalar field, respectively. Solving the Klein-Gordon equation, we find the following differential equations for the radial and angular wave function of the scalar probe
\begin{equation}
    r^2g(r)\frac{d^2R(r)}{dr^2}+(r^2\frac{dg(r)}{dr}+2rg(r))\frac{dR(r)}{dr}+V(r)R(r)=0, \label{e115}
\end{equation}
\begin{equation}
    \frac{1}{S(\theta)}\frac{d^2S(\theta)}{d\theta^2}+\frac{\cos{\theta}}{\sin{\theta}S(\theta)}\frac{dS(\theta)}{d\theta}-\frac{(a\omega+\Xi m)^2}{\sin^2{\theta}}=0,
\end{equation}
where
\begin{equation}
    V(r)=\frac{(am+\Xi\omega)^2r^2}{g(r)}.
\end{equation}

The positive roots of the metric function $g(r)$ indicates the location of the horizons of the black hole. It is necessary to expand the metric function $g(r)$ in the near-horizon region of the black hole as a quadratic polynomial in $(r-r_+)$
\begin{equation}
    g(r)\simeq K(r-r_+)(r-r_*), \label{expann}
\end{equation}
with $r_+$ the outer horizon and $K$ and $r_*$, the constants in terms of the black hole parameters
\begin{equation}
    K=\frac{1}{r_+^2}-2\Lambda,
\end{equation}
\begin{equation}
    r_*=r_+-\frac{6r_+-3M-4\Lambda r_+^3}{3-6\Lambda r_+^2}.
\end{equation}

By considering the radial equation (\ref{e115}) at the near-horizon region $\omega r_+ \ll 1$ and the limit where $\left| {{r_ + } - {r_ * }} \right| \ll {r_ + }$, the radial wave equation simplifies to
\begin{equation}
   \frac{d}{dr}[\left( {r - {r_ + }} \right)\left( {r - {r_ * }} \right)\frac{d}{dr}R\left( r \right)]+ \left[ {\left( {\frac{{{r_ + } - {r_ * }}}{{r - {r_ + }}}} \right)\mathcal{A} + \left( {\frac{{{r_ + } - {r_ * }}}{{r - {r_ * }}}} \right)\mathcal{B} + \mathcal{C}} \right]R\left( r \right) = 0, \label{e119}
\end{equation}
where the constants $\mathcal{A}$, $\mathcal{B}$ and $\mathcal{C}$ are given by
\begin{equation}
    \mathcal{A}=\frac{(r_+^2+2r_+r_*+3r_*^2)(am+\Xi\omega)^2}{K^2r_*^2(r_+-r_*)^2},
\end{equation}
\begin{equation}
    \mathcal{B}=-\frac{(3r_+^2+2r_+r_*+r_*^2)(am+\Xi\omega)^2}{K^2r_+^2(r_+-r_*)^2},
\end{equation}
\begin{equation}
    \mathcal{C}=\frac{(r_+^2+r_+r_*+r_*^2)(am+\Xi\omega)^2}{K^2r_+^2r_*^2}.
\end{equation}

Comparing the radial wave equation (\ref{e115}) and the Casimir operator given in equation (\ref{csm}), we realize that
\begin{equation}
    \mathcal{H}^2R(r)={\mathcal{\overline{H}}}^2R(r)=-\mathcal{C}R(r),
\end{equation}
from which we obtain the corresponding CFT parameters
\begin{equation}
    n_L=-\frac{1}{\sqrt{\frac{\Xi^2(r_+^2+2r_+r_*+3r_*^2)}{K^2r_*^2(r_+-r_*)^2}}(2\sqrt{\frac{r_*^2(3r_+^2+2r_+r_*+r_*^2)}{r_+^2(r_+^2+2r_+r_*+3r_*^2)}}-2)}, \label{e37}
\end{equation}
\begin{equation}
    n_R=0,
\end{equation}
\begin{equation}
    T_L=-\frac{\sqrt{\frac{r_*^2(3r_+^2+2r_+r_*+r_*^2)}{r_+^2(r_+^2+2r_+r_*+3r_*^2)}}+1}{4\pi\sqrt{\frac{a^2(r_+^2+2r_+r_*+3r_*^2)}{K^2r_*^2(r_+-r_*)^2}}(\sqrt{\frac{r_*^2(3r_+^2+2r_+r_*+r_*^2)}{r_+^2(r_+^2+2r_+r_*+3r_*^2)}}-1)},
\end{equation}
\begin{equation}
    T_R=\frac{1}{4\pi\sqrt{\frac{a^2(r_+^2+2r_+r_*+3r_*^2)}{K^2r_*^2(r_+-r_*)^2}}}. \label{e40}
\end{equation}

The behaviour of the left and right CFT temperatures are shown in figure \ref{fig:f2} for a set of values.

 \begin{figure}[h]
	\centering
	\begin{subfigure}{0.4\linewidth}
		\includegraphics[width=\linewidth]{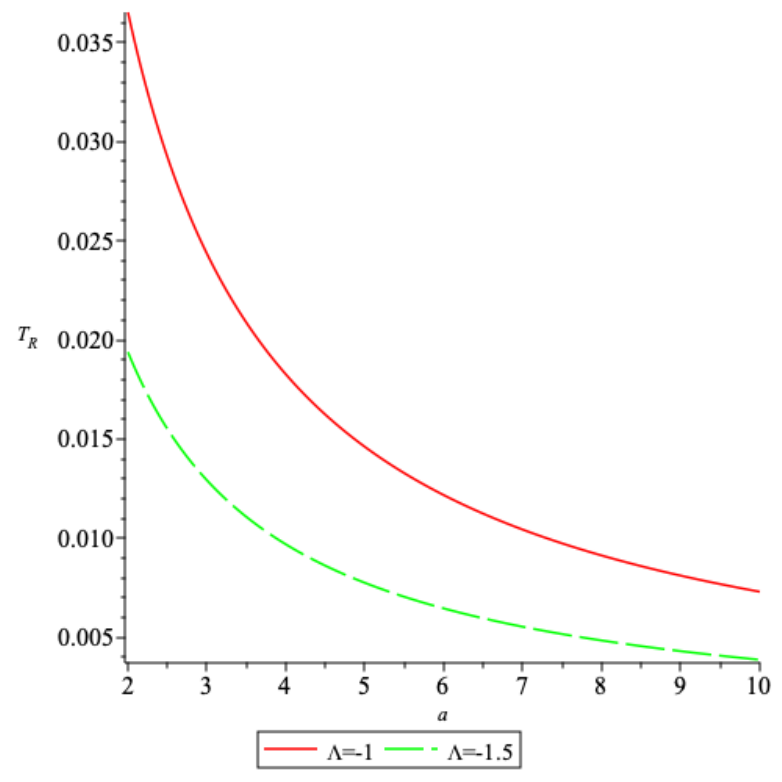}
		\caption{\centering}
		\label{fig:2a}
	\end{subfigure}
	\begin{subfigure}{0.4\linewidth}
		\includegraphics[width=\linewidth]{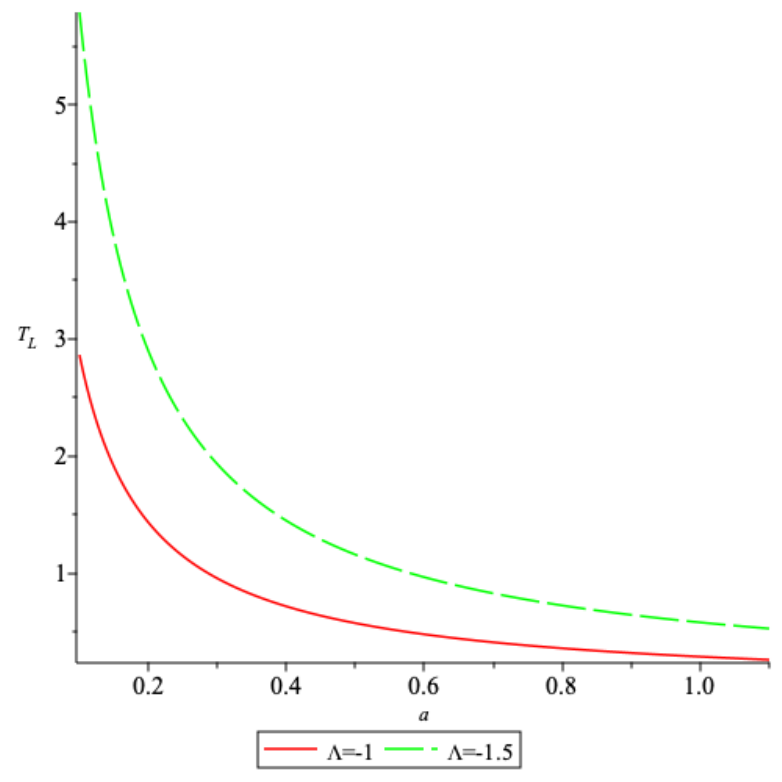}
		\caption{\centering} 
		\label{fig:2d}
	\end{subfigure}
	\caption{Behaviour of $T_{R,L}$ for $M=10$ with respect to the rotation parameter $a$}
	\label{fig:f2}
\end{figure}

We find the Bekenstein-Hawking entropy as
\begin{equation}
    S_{BH}=\frac{A}{4G}f'(R)|_{r_+}=(4/3)\pi r_+^2\Xi,
\end{equation}
which matches the microscopic Cardy entropy for the following central charges $c_L=c_R$ \cite{el2012emergent}

\begin{equation}
 c_{L}=c_R=-8r_+^2\Xi\sqrt{\frac{a^2(r_+^2+2r_+r_*+3r_*^2)}{K^2(r_+-r_*)^2r_*^2}}\Bigl(\sqrt{\frac{r_*^2(3r_+^2+2r_+r_*+r_*^2)}{r_+^2(r_+^2+2r_+r_*+3r_*^2)}}-1\Bigl).
\end{equation}
In figure \ref{fig:entbb}, we show the entropy $S_{BH}$ and central charges $c_{L,R}$ versus the rotational parameter $a$ for different values of $\Lambda$. We notice that the entropy and central charges are at their minima, where the black hole is not rotating. The entropy and the central charges are monotonic increasing functions in terms of the black hole rotational parameter.
\begin{figure}[h]
	\centering
	\begin{subfigure}{0.4\linewidth}
		\includegraphics[width=\linewidth]{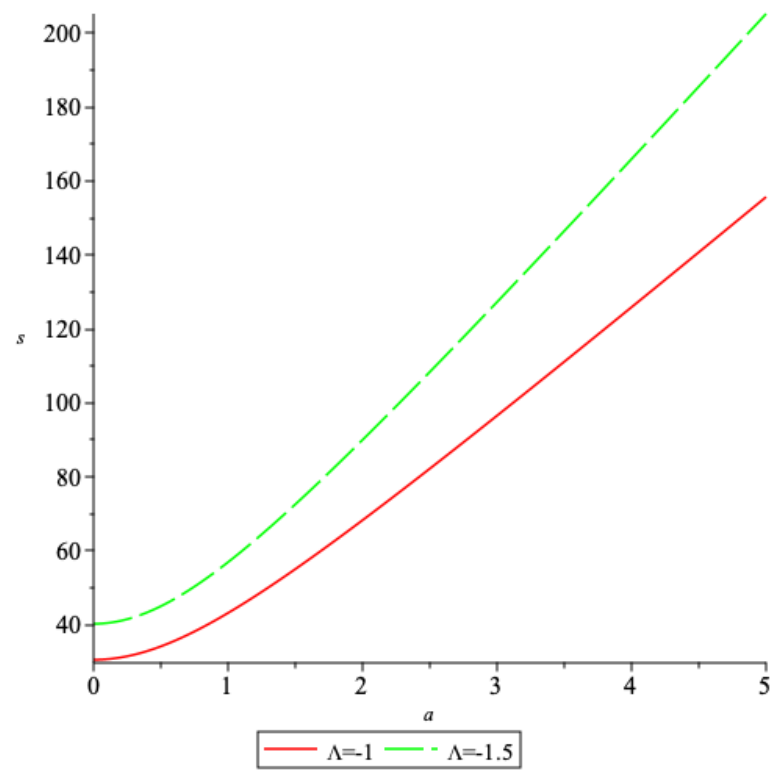}
		\caption{\centering}
		\label{}
	\end{subfigure}
	\begin{subfigure}{0.4\linewidth}
		\includegraphics[width=\linewidth]{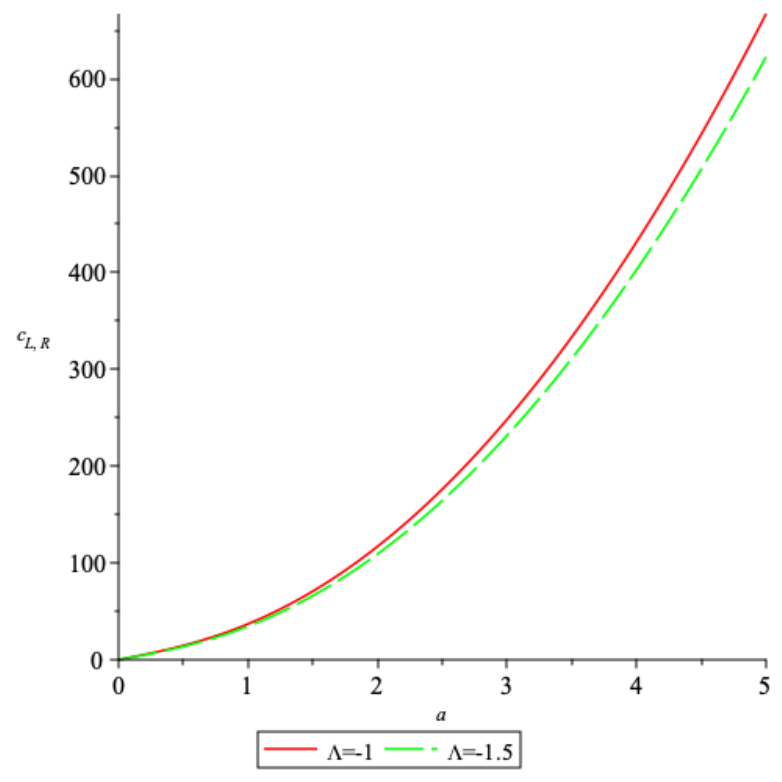}
		\caption{\centering} 
		\label{}
	\end{subfigure}
	\caption{The entropy $S_{BH}$ and central charges $c_{L,R}$ in terms of the rotational parameter $a$ for two different values of the cosmological constant $\Lambda$, where we set $M=10$. }
	\label{fig:entbb}
\end{figure}

\section{Comparison of two rotating black holes}\label{sec:sec5}
We note that the transformation (\ref{e4a}) and (\ref{e4b})  are basically a Lorentz boost in $\phi$-direction. In other words, one can always find a proper reference frame to co-rotate with the black hole. This is not possible for the first type of the rotating solution (\ref{AA metric}). We then propose a ``{\it conjecture}'' about distinguishing the {\color{black} two different} rotating black hole solutions: type I (\ref{AA metric}) and type II (\ref{e5}). To propose the ``{\it conjecture}'', we compare numerically the conformal temperatures as well as conformal mode numbers dual to the type I and type II black holes with the same mass and horizon radius. This comparison may distinguish the two different type of the rotation. To the best of our knowledge, this is the first time, that comparing CFT results for two identical rotating black holes, can be used to to distinguish between the type of rotation and presence of the matter around a black hole.
 {\textcolor{black}{\, Intuitively, for a vacuum black hole, the dual CFT states which live on the black hole horizon, should be the ground states or lower energy excited states. On the other hand, the presence of matter around the black hole solution leads to having higher energy excited states for the dual CFT on the black hole horizon. The agreement of this intuitive picture leads us to the above mentioned ``{\it conjecture''}. Of course, the main objection of the ``{\it conjecture''} is about distinguishing between two rotating black holes, one with inherent rotation, and the other with ``boosted" rotation.  The former rotating black hole has higher dual CFT temperatures and mode numbers compared to the latter black hole}.

As an example, for a set of values $M=10$ and $\Lambda=-1$ the outer horizon  $r_+$ of both types of black holes will be almost the same value $r_+=2.74$, and the two black holes are indistinguishable from the point of view of an observer far away. We plot the left and right CFT temperatures for both types of rotating black holes in figure \ref{fig:FIGT}.
We realize that the left and right CFT temperature for type I rotating black hole (no boost, in presence of matter) are higher than the corresponding temperatures for type II rotating black hole (boosted, and in vacuum).

\begin{figure}[h]
	\centering
	\begin{subfigure}{0.4\linewidth}
		\includegraphics[width=\linewidth]{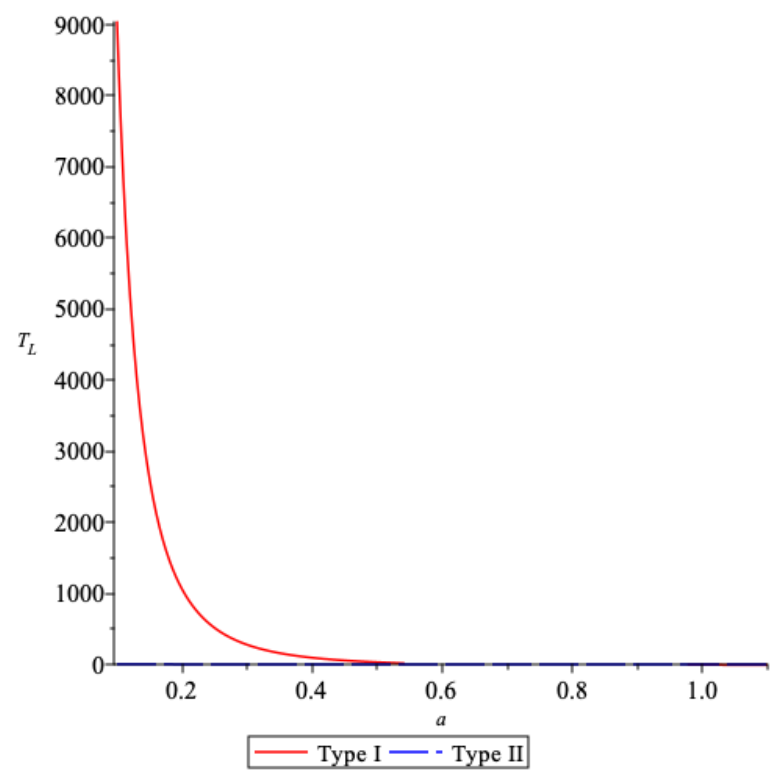}
		\caption{\centering}
		\label{fig:2a}
	\end{subfigure}
	\begin{subfigure}{0.4\linewidth}
		\includegraphics[width=\linewidth]{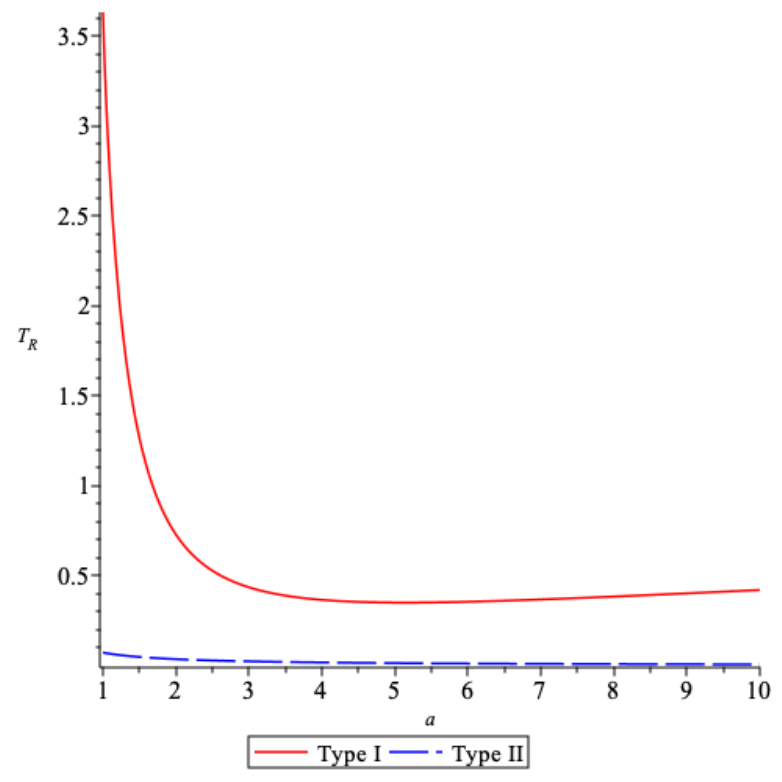}
		\caption{\centering} 
		\label{fig:2d}
	\end{subfigure}
	\caption{ Comparison of the $T_{L,R}$ for both types of rotation, when $M=10$ and $\Lambda=-1$.}
	\label{fig:FIGT}
\end{figure}

For the same values of $M=10$, $\Lambda=-1$, and  the outer horizon  $r_+=2.74$, we plot the left and right CFT mode numbers  for both types of rotating black holes in figure \ref{fig:FIGn}.
We again realize that the left and right CFT mode numbers for type I rotating black hole (no boost, in presence of matter) are higher than the corresponding mode numbers for type II rotating black hole (boosted, and in vacuum). We conclude with the following ``{\it conjecture}" that: {\it {``Two identical rotating black holes (the same mass and the horizon) in $f(R)$ theory could be distinguished by looking at their dual CFT temperatures, and the dual CFT mode numbers.  The rotating black hole in the presence of matter has higher CFT temperatures and higher CFT mode numbers, compared to the vacuum boosted rotating black hole".}}

\begin{figure}[h]
	\centering
	\begin{subfigure}{0.4\linewidth}
		\includegraphics[width=\linewidth]{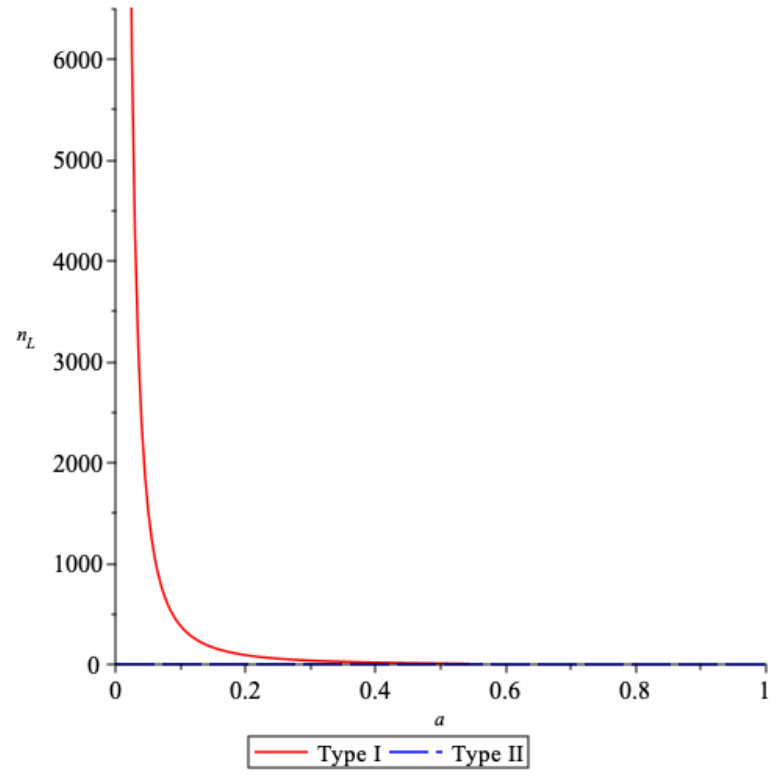}
		\caption{\centering}
		\label{fig:2a}
	\end{subfigure}
	\begin{subfigure}{0.4\linewidth}
		\includegraphics[width=\linewidth]{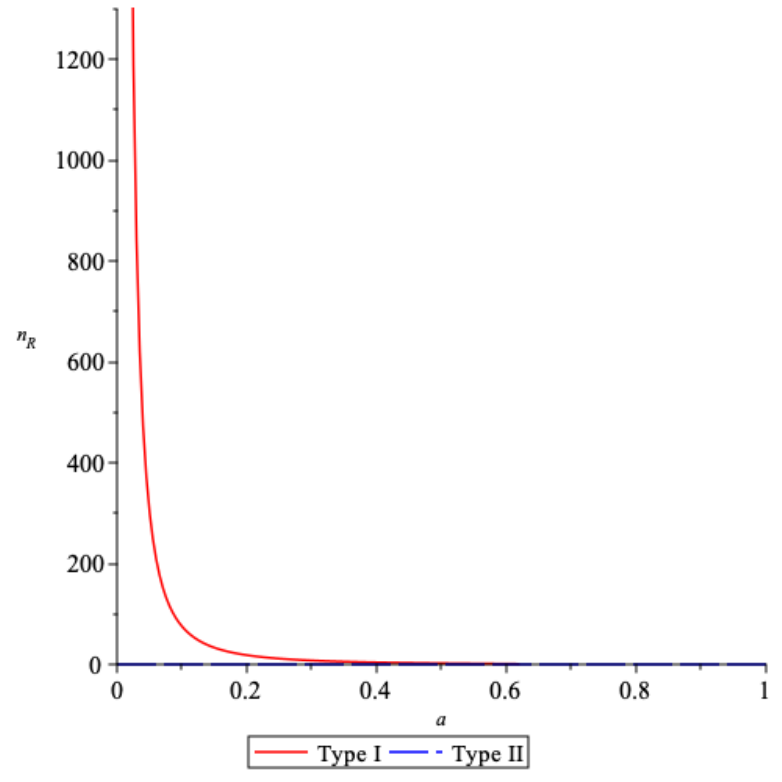}
		\caption{\centering} 
		\label{fig:2d}
	\end{subfigure}
	\caption{ Comparison of the $n_{L,R}$ for both types of rotation, when $M=10$ and $\Lambda=-1$.}
	\label{fig:FIGn}
\end{figure}

\section{Conclusions}\label{sec:conc}
{\textcolor{black}{In this article, we generalize and extend the very well known black hole holography in GR, to two very different rotating black holes in the $f(R)$ gravity. }}

{\textcolor{black}{We explicitly construct a rotating black holes in $f(R)$ theory using a modification of the Newman-Janis algorithm. We then numerically verify that the existing non-zero components of the energy-momentum tensor satisfy the three energy conditions: the weak, strong and dominant energy conditions.  We use the field equation of a probe field in the background of rotating black hole, to establish the hidden conformal symmetry for the rotating black holes in $f(R)$ theory.  
}}

{\textcolor{black}{To achieve the establishing the black hole holography, we mainly consider the near-horizon region, since the line element metric function, which determines the location of the event horizon, is a quartic algebraic equation. In the near-horizon region, we show that the radial part of the field equation for the scalar probe field, could be {\color{black} compared} to the quadratic Casimir equation for an $SL(2,\mathbb{R})_L\times SL(2,\mathbb{R})_R$ conformal algebra. The existence of conformal algebra shows a local hidden conformal symmetry which acts on the solution space of scalar probe field. }}

{\textcolor{black}{We also explicitly show that the $SL(2,\mathbb{R})_L\times SL(2,\mathbb{R})_R$ symmetry is spontaneously broken to $U(1)_L \times U(1)_R$ under the coordinate identification $\phi \sim \phi + 2\pi$. The  $U(1)_L \times U(1)_R$ subalgebra of the original conformal symmetry indicates that the rotating black holes in $f(R)$ gravity, is dual to the finite temperatures $(T_L,T_R)$ mixed state in a two-dimensional CFT. }} 

{\textcolor{black}{We also calculate the central charges of the CFT, by {\color{black} comparing} the dual CFT Cardy entropy to the macroscopic Bekenstein-Hawking entropy of the black hole.  The results very clearly indicate that the 
rotating black holes in $f(R)$ theory with particular values of black hole parameters, are indeed dual to the two-dimensional CFT.}}

{\textcolor{black}{We then consider the class of rotating black holes in the $f(R)$ gravity, obtained through using a pair of Lorentz boost transformation to the static spherically symmetric solution. Using a new set of conformal coordinates, we {\color{black} compare} the radial part of the field equation, for the scalar probe field, to the quadratic Casimir invariant of the conformal symmetry. The results of the article, for the first time, show the validity and extension of the holographic duality for two different rotating black holes, beyond the GR black holes, and to the realm of the  $f(R)$ theories of gravity.}

We also find the interesting result that we suggest as a {\it conjecture} that, 
{\it {``Two identical rotating black holes (the same mass and the horizon) in $f(R)$ theory could be distinguished by looking at their dual CFT temperatures, and the dual CFT mode numbers.  The rotating black hole in the presence of matter has higher CFT temperatures and higher CFT mode numbers, compared to the vacuum boosted rotating black hole".}}

{\textcolor{black}{For future works, we can construct the extended conformal symmetry for the possible charged rotating black holes in $f(R)$ gravity. The charged rotating black holes have two dual CFTs which we refer to them as the $J$ and $Q$ pictures. In the extended conformal symmetry, the underlying symmetry is deformed by a deformation parameter \cite{NN2}. We also plan to verify or reject the existence of a super conformal field theory, with a global $U(1)$ symmetry, dual to charged rotating black holes in $f(R)$ gravity. The dual $J$ and $Q$ pictures are related to the spectral flow transformations of the super conformal field theory. 
Finding the new charged rotating black hole solutions in MTG theories, such as $f(R)$ and {\color{black} $f({\bold Q})$ (${\bold Q}$ is related to the non-metricity of metric)} theories, not only provides a huge amount of information about the charged rotating black holes in MTG, but also may provide establishing (or rejecting) the black hole holography,  in $f(R)$ and {\color{black} $f({\bold Q})$} theories.  The establishments of the holography for charged rotating black holes beyond GR, may shed lights on the nature of the black hole holography in the context of quantum gravity. 
}}

{\textcolor{black}{The other interesting open question is finding the near-horizon geometry of the near-extremal charged rotating black holes of the $f(R)$ and {\color{black} $f({\bold Q})$} theories.  
Finding the central charges of the dual CFT to the charged rotating solutions of the MTG, using the asymptotic symmetry group, is another interesting open question. Though it is a very complicated task, however it provides another layer of confirmation on the duality between charged rotating black holes of MTG and CFTs, and especially our results in this article for the rotating black holes of $f(R)$ gravity.  
Finding the different types of the super-radiant scattering, off the near-extremal black holes of MTG is another interesting research project.  The results definitely provide further evidence to support the holographic dual for the charged rotating black holes of MTG, as well as the charged rotating black holes of the $f(R)$ gravity. Moreover studying the proposed ``{\it conjecture}'' in this article, in other $f(R)$, $f(T)$ {\color{black} (${T}$, the torsion scalar is related to the torsion tensor)} and {\color{black} $f({\bold Q})$}  theories may provide {\color{black} a universal} feature for the ``{\it conjecture}''.
}}

\section{Appendix A} \label{App:A}
In this appendix, we list all non-zero components of the energy momentum tensor for the imperfect fluid. 

\begin{eqnarray}
    T_{tt} &=& -\frac{6a^2}{r^6(r^2+a^2\cos^2{\theta})^3}\Bigl[-\frac{8}{9}a^6\cos^8{\theta}(-1/24\Lambda r^4+Mr-5/4a^2+1/2r^2)+a^4\cos^6{\theta}
\nonumber \\
&\times& \Bigl(11/18\Lambda r^5+1/3\Lambda Mr^4-4/3r^3-11/27\Lambda a^2r^3-5/3Mr^2+32/9a^2r+M^2r
\nonumber \\
&-& 11/9Ma^2\Bigl) +4/3r^3a^2\cos^4{\theta}\Bigl(-7/72\Lambda^2r^7+17/24\Lambda r^5+1/24\Lambda Mr^4-2/3a^2\Lambda r^3-r^3
 \nonumber \\
 &-& 1/2Mr^2+M^2r+3a^2r-2Ma^2\Bigl)+1/3r^5\cos^2{\theta}\Bigl(-5/9\Lambda^2r^7+19/9\Lambda r^5-4/3\Lambda Mr^4
 \nonumber \\
 &-& 2/3a^2\Lambda r^3-4/3r^3+1/3Mr^2+M^2r+16/3a^2r-5Ma^2\Bigl)-2/9r^7
 \nonumber \\
 &\times& (M-r+1/3\Lambda r^3)\Bigl],
\end{eqnarray}

\begin{eqnarray}
   T_{rr} &=& -\frac{2a^2}{3(r^2+a^2\cos^2{\theta})(1/3\Lambda r^4+Mr-a^2-r^2)r^5}\Bigl[a^4(10/3\Lambda r^3+M-10r)\cos^6{\theta}
    \nonumber \\
    &+& 4a^2r\cos^4{\theta}(53/24\Lambda r^4+Mr+1/2a^2-5r^2)+\cos^2{\theta}\Bigl(7\Lambda r^7+3Mr^4+4a^2r^3
    \nonumber \\
    &-& 10r^5\Bigl)+2r^5\Bigl],
\end{eqnarray}

\begin{eqnarray}
    T_{\theta\theta} &=& -\frac{16a^2\cos^2{\theta}}{3r^6(r^2+a^2\cos^2{\theta})}\Bigl[a^4\cos^4{\theta}(-1/24\Lambda r^4 +Mr-5/4a^2-1/4r^2)
    \nonumber \\
    &+& 7/4a^2r^2\cos^2{\theta}(-17/84\Lambda r^4+Mr-10/7a^2-2/7r^2)-1/8\Lambda r^8+3/4Mr^5
    \nonumber \\
    &-& 5/4a^2r^4-1/4r^6\Bigl],
\end{eqnarray}

\begin{eqnarray}
    T_{\phi\phi} &=& \frac{6a^2\sin^2{\theta}}{r^6(r^2+a^2\cos^2{\theta})^3}\Bigl[a^6\cos^8{\theta}(Mr-10/9a^2)(1/3\Lambda r^4+Mr-a^2-r^2)
    \nonumber \\
    &-& a^4r\cos^6{\theta}\Bigl(-7/54\Lambda^2r^9-5/18\Lambda r^7-1/18\Lambda Mr^6+1/2a^2\Lambda r^5+1/3M\Lambda a^2r^4
    \nonumber \\
    &+& 8/3Mr^4-11/27a^4r^3\Lambda -4/3M^2r^3-32/9a^2r^3+5Ma^2r^2+M^2a^2r-32/9a^4r
    \nonumber \\
    &-& 11/9Ma^4\Bigl) -4/3a^2r^3\cos^4{\theta}\Bigl(5/36\Lambda^2r^9-7/72a^2\Lambda^2r^7-13/24\Lambda r^7+1/3\Lambda Mr^6
    \nonumber \\
    &-& 5/24a^2\Lambda r^5+1/24Mr^4a^2\Lambda +21/12Mr^4-2/3a^4\Lambda r^3-1/4M^2r^3-3a^2r^3
    \nonumber \\
    &+& 11/4Ma^2r^2+M^2a^2r-3a^4r-2Ma^4\Bigl)-1/3r^5\cos^2{\theta}\Bigl(-5/9a^2\Lambda^2r^7-1/3\Lambda r^7
    \nonumber \\
    &+& 7/9a^2\Lambda r^5-4/3\Lambda Ma^2r^4 +2Mr^4-2/3a^4\Lambda r^3-16/3a^2r^3+7/3Ma^2r^2+M^2a^2r
    \nonumber \\
    &-& 16/3a^4r-5Ma^4\Bigl)+2/9r^7(1/3a^2\Lambda r^3+r^3+a^2r+Ma^2)\Bigl],
\end{eqnarray}

\begin{equation}
T_{r\theta}=\frac{-4a^2\cos{\theta}\sin{\theta}(r^2+a^2\cos^2{\theta})}{r^5},
\end{equation}

\begin{eqnarray}
    T_{t\phi} &=& \frac{6a^3\sin^2{\theta}}{r^5(r^2+a^2\cos^2{\theta})^3}\Bigl[a^4\cos^6{\theta}(1/3\Lambda r^3+M)(Mr-11/9a^2+1/3r^2)
    \nonumber \\
    &+& 4/3a^2r^2\cos^4{\theta}(1/3\Lambda r^3+M)(-7/24\Lambda r^4+Mr-2a^2+1/2r^2)+1/3r^4\cos^2{\theta}
    \nonumber \\
    &\times& (-5/9\Lambda^2r^7-4/3\Lambda Mr^4-2/3r^3a^2\Lambda +4/3\Lambda r^5+M^2r-5Ma^2+Mr^2)
    \nonumber \\
    &-& 2/9r^6(1/3\Lambda r^3+M)\Bigl].
\end{eqnarray}

\section{Appendix B}

The radial part of the Klein-Gordon equation can be written as
\begin{equation}
   \frac{d}{dr}[\left( {r - {r_ + }} \right)\left( {r - {r_ * }} \right)\frac{d}{dr}R\left( r \right)]+ \left[ {\left( {\frac{{{r_ + } - {r_ * }}}{{r - {r_ + }}}} \right)\mathcal{A} + \left( {\frac{{{r_ + } - {r_ * }}}{{r - {r_ * }}}} \right)\mathcal{B} + \mathcal{C}} \right]R\left( r \right) = 0, \label{e192}
\end{equation}
where $\mathcal{A}$, $\mathcal{B}$ and $\mathcal{C}$ are constants and given by
\begin{eqnarray}
    \mathcal{A} &=& -\frac{2r_*}{(r_+-r_*)^2(Kr_+^3r_*+a^2)^3Kr_+}\Bigl(-1/2K^2\omega^2r_+^{12}-K^2\omega^2r_+^{11}r_*-K^2\omega r_+^{10}(3/2r_*^2\omega
    \nonumber \\
    &+& a^2\omega -ma)+ K^2ar_*\omega(a\omega-m)(Kr_*^2-2)r_+^9 -1/2K^2a(a\omega-m)(a^2\omega+6r_*^2\omega
    \nonumber \\
    &-& ma)r_+^8-a^2r_*Kr_+^7(Ka^2\omega^2-2Kam\omega+Km^2+3/2\omega^2)+r_+^6(3/2K^2a^4r_*^2\omega^2
    \nonumber \\
    &-& 3/2K^2a^2m^2r_*^2)-3Ka^3r_*\omega r_+^5(a\omega-m)-1/2a^4\omega^2r_+^4+(3/2Ka^6r_*\omega^2
    \nonumber \\
    &-& 3/2Ka^4m^2r_*)r_+^3+r_+^2(a^5m\omega-a^6\omega^2)+1/2a^8\omega^2-1/2a^6m^2\Bigl),
\end{eqnarray}
\begin{eqnarray}
    \mathcal{B} &=& -\frac{2r_*}{(r_+-r_*)^2(Kr_+^3r_*+a^2)^3Kr_+}\Bigl(-3/2K^2\omega^2r_+^{10}r_*^2+K^2r_*^3\omega r_+^9(Ka^2\omega-Kam-\omega)
    \nonumber \\
    &-& 3r_*^2\omega K^2r_+^8(1/6r_*^2\omega+a^2\omega-am)-2ar_*Kr_+^7\omega(K(a\omega-m)r_*^2+3/4a\omega) 
    \nonumber \\
    &+& 3/2ar_*^2(a\omega-m)K^2r_+^6(-2/3r_*^2\omega+a(a\omega+m))-a^2r_*(K(a\omega-m)r_*^2
    \nonumber \\
    &+& +3a\omega)(a\omega-m)Kr_+^5-1/2a^2(K^2(a\omega-m)^2r_*^4+a^2\omega^2)r_+^4+r_*r_+^3(3/2Ka^6\omega^2
    \nonumber \\
    &-& 3/2Ka^4m^2)+(a^5m\omega-a^6\omega^2)r_+^2+1/2a^8\omega^2-1/2a^6m^2\Bigl),
\end{eqnarray}
\begin{equation}
    \mathcal{C}=\frac{r_+r_*(a^2\omega+\omega r_+^2-am)^2(Kr_+^4+Kr_+^3r_*+Kr_+^2r_*^2-a^2)}{(Kr_+^3r_*+a^2)^3}.
\end{equation}

\bigskip

{\Large Acknowledgments}\newline
This work was supported by the Natural Sciences and Engineering Research Council of Canada.\newline

\end{document}